\documentclass[12pt]{article}
\usepackage{amsfonts}
\usepackage{amsmath}
\usepackage{amssymb}
\usepackage{amscd}
\usepackage{epsfig}
\usepackage{subeqnarray}
\usepackage{array}
\usepackage{afterpage}
\usepackage{hhline}

\allowdisplaybreaks[4]

\def\p{{|{\bf p}|}}
\def\q{{|{\bf q}|}}
\def\S{\mathfrak{S}}
\def\T{\mathfrak{T}}

\def\N{\mathfrak{N}}
\def\D{\mathfrak{D}}
\def\Q{\mathfrak{Q}}
\def\rb#1{\raisebox{8pt}[26pt][0pt]{#1}}
\def\Not#1{\not\hskip-3pt#1}
\def\intq#1{\int \frac{d^4#1}{(2\pi)^4}}
\def\intt#1{\int \frac{d^3#1}{(2\pi)^3}}
\def\Tr{{\rm Tr}}

\begin{document}

\title{The generalised relativistic Lindhard
  functions}

\author{M. B. Barbaro$^1$, R. Cenni$^2$
and M. R. Quaglia$^2$\\
${}^1$ Dipartimento di Fisica Teorica --- Universit\`a di Torino
\\
Istituto Nazionale di Fisica Nucleare --- Sez. di Torino\\
 Torino ---Italy \\
${}^2$ Dipartimento di Fisica --- Universit\`a di Genova\\
Istituto Nazionale di Fisica Nucleare --- Sez. di Genova\\
Genova --- Italy\\}
\date{}

\maketitle
\begin{abstract}
We present here analytic expressions for the generalised Lindhard
function, also referred to as Fermi Gas polarisation
propagator, in a relativistic kinematic framework and
in the presence of various resonances and vertices.
Particular attention is payed to its
real part, since it gives rise to substantial difficulties in the definition
of the currents entering the dynamics.
\end{abstract}

\section{Introduction}
\label{sec:1}

The linear response theory, whose key ingredient is the Lindhard
function (LF), is usually formulated in many-body frameworks like RPA, 
Landau quasi-particle theory and the like (see, {\it e.g.},
Ref.~\cite{FeWa-71-B}). 
Noteworthy, these simple approaches are often able
to describe successfully a quite involved physics. 

The LF~\cite{Li-54}, 
as originally defined, is just the particle-hole polarisation propagator
for a non-relativistic free Fermi gas (FFG) of electrons, and reads
\begin{equation}
  \label{eq:001}
  \Pi^0(q_0,\q)=-2i\intq{p}G_0(p+q)G_0(p)~,
\end{equation}
where 
\begin{equation}
  \label{eq:002}
  G_0(p)=\frac{\theta(\p-k_F)}{p_0-\dfrac{\p^2}{2m}+i\eta}+
  \frac{\theta(k_F-\p)}{p_0-\dfrac{\p^2}{2m}-i\eta}
\end{equation}
is the Green's function of the free electron.
The factor 2 in front of \eqref{eq:001} comes
from the spin traces and is replaced by a factor 4 in nuclear physics
(spin plus isospin).

The explicit form of \eqref{eq:001} is known since 1954~\cite{Li-54} and,
more recently~\cite{CeSa-88,CeCoCoSa-92},
has been expressed according to
\begin{equation}
  \label{eq:003}
  \Pi^0(q_0,\q)=2m\left[I[y(\q,q_0+i\eta{\rm \ sgn} q_0),\q]+
    I[y(\q,-q_0-i\eta{\rm \ sgn} q_0),\q]\right]\;,
\end{equation}
where 
\begin{equation}
  \label{eq:006}
   I[y,\q]=\frac{1}{(2\pi)^2\q}\left\{yk_F+\frac{1}{2}
    (y^2-k_F^2)\log\frac{y-k_F}{y+k_F}\right\}~,
\end{equation}
\begin{equation}
  \label{eq:004}
  y(\q,q_0)=\frac{m q_0}{\q}-\frac{\q}{2}
\end{equation}
being the West's scaling variable~\cite{We-75}. The analytic extension when 
the arguments of the logarithms become negative is prescribed to be
\begin{equation}
  \label{eq:A501}
  \log(x\pm i\eta)=\log|x|\pm i\pi\theta(-x)
\end{equation}
and the imaginary part
of the LF, namely the response to a scalar(-isoscalar) probe,
is thus obtained.

Actually many calculations of a fermionic relativistic response 
function are available 
--- and hence a number of relativistic many-body computations have
been performed. Still the need of analytic
expressions for the relativistic generalisation of the LF, 
as an useful input for a variety of calculations, is felt: indeed the
results presently available (see, {\it e.g.}, Ref.~\cite{Fa-98},
which completes a previous result~\cite{KuSu-85}, and Ref.~\cite{Am-al-02}) 
mostly refer to the electroweak case. However we are also interested to the 
relativistic response to pions and $\rho$-mesons, which lead to
different generalisations of the LF. Further, the 
excitation of nucleonic resonances need to be accounted for and, last but not
least, in the quark-gluon plasma the case of massless particles (or one massive
--- an $s$ or a $c$ quark --- and one massless) deserves some attention.

In this paper we address a number of the above cases, showing 
that they can be handled 
algebraically in terms of only a few explicit functions.

The scope we pursue is to provide, as a hopefully useful tool
for people involved in the field, a rather comprehensive description
of the generalised LF in the relativistic case for 
nucleons and for 1/2 and 3/2 spin resonances.
The limiting case quoted above will also be examined in detail.

\section{Setting the problem}

\label{sec:2}

The channel dependence of the FFG polarisation propagator in a relativistic
framework is more pronounced than in the non-relativistic case.
We start with the generalised  LF for a non interacting nucleonic system
\begin{equation}
  \label{eq:C001}
  \Pi^0_{xy}(\q,q_0)=-i\Tr\intq{p}S_m(p+q){\cal O}^xS_m(p){\cal O}^y~,
\end{equation}
where
\begin{equation}
  \label{eq:C002}
  \begin{split}
    S_m(p)&=\frac{\Not p+m}{2E_p}
    \left\{\frac{\theta(|{\bf p}|-k_F)}{p_0-E_p+i\eta}
      +\frac{\theta(k_F-|{\bf p}|)}{p_0-E_p-i\eta}-\frac{1}{p_0+E_p-i\eta}
    \right\}\\
    &=(\Not p+m)\S_m(p)\\
    &=S_0(p)+\frac{\Not p+m}{2E_p}2i\pi\delta(p_0-E_p)\theta(k_F-|{\bf p}|)
  \end{split}
\end{equation}
is the nucleon [electron, quark...] propagator in the medium, with
\begin{equation}
  \label{eq:C003}
  E_p=\sqrt{p^2+m^2}~,
\end{equation}
the indices $x$ and $y$ label the incoming and outgoing channel 
(not necessarily coincident) and
${\cal O}^x$, ${\cal O}^y$ are some combinations of $\gamma$ matrices 
and momenta embodying the vertices characterising the channels. 
Moreover, $S_0$ is the analogous of $S_m$ in the vacuum.

We have also introduced a ``reduced''
fermion propagator $\S_m(p)$ spoiled of the Dirac matrix structure, its
inverse reading
\begin{equation}
  \label{eq:A480}
  \D(p)=p^2-m^2~.
\end{equation}

Although $\Pi^0_{xy}$ as given in Eq.~\eqref{eq:C001} is ill-defined since it
is expressed by a divergent integral,
we will not require renormalisability, since one may
also be interested  in effective theories. However, a regularisation
procedure is needed in order to cancel the divergences. 
In the case of \eqref{eq:C001}
(one-loop level) the vacuum subtraction is sufficient~\cite{AlCeMoSa-88}. 

Thus, defining the following polynomial in the 
relativistic invariants $p^2$, $p\cdot q$ and $q^2$
\begin{equation}
  \label{eq:C005}
  f_{xy}(p,q)=\Tr(\Not p+m){\cal O}^x(\Not p+\Not q+m){\cal O}^y~,
\end{equation}
$\Pi^0_{xy}$ will read
\begin{equation}
  \label{eq:C004}
  \Pi^0_{xy}(\q,q_0)=-i\intq{p}
  \left[\S_m(p+q)\S_m(p)-\S_0(p+q)\S_0(p)\right]f_{xy}(p,q)~,
\end{equation}
where $\S_0=(\Not p+m)^{-1} S_0$.

The frequency integration in \eqref{eq:C004} then reduces to the
evaluation of the residua in the (say) lower half-plane, because along the
half-circle at infinity the $\pm i\eta$
in the denominators become irrelevant and the  integrand  vanishes. 
Thus the regularised $\Pi^0_{xy}$ is given by
\begin{equation}
  \begin{split}
    \Pi^0_{xy}&(\q,q_0)=
    \intt{p}\frac{1}{4E_pE_{p+q}}
    \label{eq:C006}
    \\
    &\times\left\{ \frac{\theta(|{\bf p+q}|-k_F)\theta(k_F-|{\bf p}|)}
        {q_0-E_{p+q}+E_p+i\eta}f_{xy}(p,q)\bigm|_{p_0=E_{p+q}-q_0}   \right.  
    \\
    &-
      \frac{\theta(k_F-|{\bf p+q}|)\theta(|{\bf p}|-k_F)}
        {q_0-E_{p+q}+E_p-i\eta}f_{xy}(p,q)\bigm|_{p_0=E_p}          
    \\
    &+
    \frac{\theta(k_F-|{\bf p+q}|)}
         {q_0-E_{p+q}-E_p+i\eta}f_{xy}(p,q)\bigm|_{p_0=E_{p+q}-q_0}  \\-
    &\frac{\theta(k_F-\p)}{q_0+E_{p+q}+E_p-i\eta}f_{xy}(p,q)\bigm|_{p_0=E_p}\\
    &+\theta(|{\bf p+q}|-k_F)\theta(|{\bf p}|-k_F)
    \left.
      \frac{f_{xy}(p,q)\bigm|_{p_0=E_{p+q}-q_0}-f_{xy}(p,q)\bigm|_{p_0=E_p}}
    {q_0-E_{p+q}+E_p}
    \right\}
  \end{split}
\end{equation} 
(note that the term in the last line is never singular).

Now in each denominator  the factor $\pm i\eta$
can be replaced by $+i\eta{\rm \ sgn}(q_0)$. 
In fact the denominators in the first
and third term can only vanish when $q_0>0$ so here the replacement
$i\eta\to i\eta{\rm \ sgn}(q_0)$ is immaterial, while the 
second and fourth ones can vanish for $q_0<0$ so that the term $-i\eta$
plays the same role of $i\eta{\rm \ sgn}(q_0)$.

Since, as we shall see, the explicit form of $f_{xy}$
is inessential for the following discussion, 
we consider the case $f_{xy}(p,q)=1$.
Then, with some manipulations of the $\theta$ functions and a change of 
variable, Eq.~\eqref{eq:C006} simplifies to
\begin{equation}
  \label{eq:C016}
    \begin{split}
      \Pi^0(\q,q_0)&=\intt{p}\frac{\theta(k_F-|{\bf p}|)}{2E_p}\\
      &\hskip-20pt\Biggl\{\frac{1}{(q_0+i\eta{\rm \ sgn}(q_0)+E_p)^2-E_{p+q}^2}
      +\frac{1}{(q_0+i\eta{\rm \ sgn}(q_0)-E_p)^2-E_{p+q}^2}\Biggr\}~.
\end{split}
\end{equation}
Eq.~\eqref{eq:C016}  displays a great advantage
from a practical point of view since 
\begin{enumerate}
\item each term contains only one $\theta$ function, hence
the analytic calculation of the integrals is simplified;
\item  $\Pi^0$ can be evaluated in a region
  where it is real and then its imaginary part follows by analytic
  extension by suitably 
  approaching the real axis in the complex plane of $q_0$;
\item it is manifestly even in $q_0$.
\end{enumerate}
The same procedure led to the form \eqref{eq:003}
for the non-relativistic LF.

Now consider a $N^*$ excitation of mass $M$.
In this case the polarisation propagator at the lowest order is 
built up by {\em two}
non-coincident Feynman diagrams, namely 
(the star will always denote quantities involving a
resonance or a resonance-hole pair)
\begin{multline}
  \label{eq:C007}
  \Pi^{*}_{xy}(\q,q_0)=-i\Biggl[\intq{p}\S^*(p+q)\S_m(p)f^{*xy}(p,q)\\
  +\intq{p}\S^*(p-q)\S_m(p)f^{*xy}(p,-q)\Biggr]~,
\end{multline}
where 
\begin{equation}
  \label{eq:C008}
  \S^*(p)=\left[{p^2-M^2+i\eta}\right]^{-1}~.
\end{equation}

For later purposes we also introduce the inverse of  $\S^*$, namely
\begin{equation}
  \label{eq:A473}
  \D^*(p)=p^2-M^2~.
\end{equation}
The explicit form of $f^{*xy}$  will be specified later.

Clearly, each term in \eqref{eq:C007} contains only one 
$\theta$-function because
there is no Pauli blocking. Convergence
is ensured by vacuum subtraction and, at variance of the nucleon-hole case,
Eq.~\eqref{eq:C002} (third line) tells us that $p_0$ can 
be replaced everywhere by $E_p$, since $S_0$ never contributes.

\section{Structure of the Lindhard functions}
\label{sec:3}

In this Section we set up the general structure of the 
relativistic LFs, which will be later evaluated in some
specific cases.

Before presenting the detailed calculation, we observe
that the Lorentz covariance is broken by the presence
of an infinite medium like the FFG, 
since this naturally selects a privileged frame
of reference, namely the one in which the FFG is at rest.
Indeed in this system the nuclear matter has zero momentum, while
any boost, no matter how small the velocity is, generates a state with
infinite momentum. 

If we instead consider a system with mass $M$ and finite momentum 
${\bf p}$, then any response function $f$ to a probe carrying a four-momentum 
$q_\mu=(q_0,{\bf q})$ can only depend upon Lorentz scalars, namely
$f=f(p^2=M^2,p\cdot q,q^2)$; however in the $M\to\infty$ limit 
$p\cdot q=q_0\sqrt{p^2+M^2}-{\bf p\cdot q}\to q_0 M$, so that the 
Lorentz covariance is broken since $f$ will depend upon $q_0$ and 
${\bf q}$ {\em separately}.

\subsection{The ingredients}
\label{sec:3.1}

Having clarified the functional dependence of the LFs,
we now introduce the ingredients needed to their evaluation.
We define the functions
\begin{equation}
  \label{eq:C009}
  U^{*[n]}_{rel}(M;\q,q_0)=\intt{p}\frac{E^n_p}{2E_p}
  \frac{\theta(k_F-\p)}{(p+q)^2-M^2+i\eta}\Biggm|_{p_0=E_p}~,
\end{equation}
to be computed in the next Section.
The quantities of direct physical interest are the even and odd parts
(in $q_0$) of $U^{*[n]}_{rel}$, namely
\begin{equation}
  \label{eq:C109}
  \Upsilon^{*[n]}_\pm(M;\q,q_0)=U^{*[n]}_{rel}(M;\q,q_0)
  \pm U^{*[n]}_{rel}(M;\q,-q_0)~.
\end{equation}
We shall consider in the following a variety of functions $f_{xy}$, each
one giving rise to its own LF: remarkably
these all are expressed in terms of few basic cases.
Note that when $f_{xy}=1$ then Eq.~\eqref{eq:C016} becomes
\begin{equation}
\Pi^0(\q,q_0)=\Upsilon^{[0]}_+(m;\q,q_0+i{\rm \ sgn}q_0)~.
\end{equation}
The above functions display a complex analytic structure 
(logarithmic cuts)
and have a well defined imaginary part and  a well
defined asymptotic behaviour (in $q_0$), namely 
$\Upsilon^{*[n]}_\pm\simeq q_0^{-2}$: this means that the real part of 
$\Upsilon^{*[n]}_\pm$, hence of the LF, 
can be univoquely recovered from its imaginary part via dispersion relations.

However in general $f_{xy}$ is a polynomial in $q_0^2$. 
Each term of this polynomial generates contributions to $\Pi^0$ with the same
imaginary part (up to trivial coefficients), but with different
asymptotic behaviour, so that the evaluation of the real part will require
subtracted dispersion relations. 
The subtracted parts will be called contact terms. 
These can be expressed in terms of the $k_F$-dependent function
\begin{align}
  \label{eq:A422}
  \begin{split}   
      {\mathfrak T}^{[0]}=-i\intq{p}\S_m(p)&
      \begin{CD}
        @>{\text{renormalisation}}>>
    \end{CD}
    \intt{p}
      \dfrac{\theta
        (k_F-\p)}{2E_p}
    \\
    &=\frac{1}{8\pi^2}\left[k_FE_F-m^2 \log\frac{k_F+E_F}{m}\right] ~, 
  \end{split}
\end{align}
where  $E_F=\sqrt{k_F^2+m^2}$ and, more generally,
\begin{equation}
  \label{eq:A637}
  {\mathfrak T}^{[n]}=\intt{p}
      \dfrac{\theta
        (k_F-\p)}{2}E_p^{n-1}=
      \frac{k_F^3}{12\pi^2} m^{n-1} 
      {}_2F_1\left(\frac{3}{2},\frac{1}{2}-\frac{n}{2};\frac{5}{2};-
        \frac{k_F^2}{m^2}\right)~,
\end{equation}
where ${}_2F_1$ is an hypergeometric function.

Note that contact terms are also indirectly 
related to the renormalizability of a theory, because their
presence alter the power counting in a bosonic loop entering in an
RPA-dressed bosonic propagator closed on itself or on a fermionic line.

Now we are in a position to deal with the general structure of the LFs. 
Since we shall consider (pseudo-)scalar and (pseudo-)vector couplings,
our LFs will carry 0, 1 or 2 vector indices only.
Tensor couplings could bring into play other functions
($ U^{*[3]}_{rel}$ and $ U^{*[4]}_{rel}$), but they seem not to be, at present,
of physical interest.

\subsection{Scalar case: 0-index functions}

\label{sec:3.2}

Here ${\cal O}^x$ and ${\cal O}^y$ have no vector structure, hence
$f^*_{xy}=f^*_{xy}(p^2,p\cdot q,q^2)$, in general a polynomial, 
is a Lorentz scalar.
Replacing then $p^2$ with $\D(p)+m^2$ and using the identity
\begin{equation}
  \label{eq:A710}
  \begin{split}
    p\cdot q&=\frac{1}{2}
    \left[(p+q)^2-M^2\right]-\frac{1}{2}(p^2-m^2)-
    \frac{1}{2}\rho q^2\\
    &= \frac{1}{2}\D^*(p+q)-\frac{1}{2}\D(p)-\frac{1}{2}\rho q^2~,
  \end{split}
\end{equation}
we can rewrite $f^*_{xy}$ in the form
\begin{equation}
  \label{eq:A860}
  f^*_{xy}=A_{xy}^S(q^2)+\sum_{mn}\lambda_{mn}^S(q^2)
  \left\{\D^*(p+q)\right\}^m
  \left\{\D(p)\right\}^n~,
\end{equation}
being $A^S(q^2)$ and $\lambda_{mn}^S(q^2)$'s Lorentz scalars
(hence the superscript $S$).
In  \eqref{eq:A710} we have introduced, as in \cite{Am-al-99},
the dimensionless quantity
\begin{equation}
  \label{eq:AC315}
  \rho=1-\frac{M^2-m^2}{q^2}~.
\end{equation}
In the case $M=m$ we have $f^*_{xy}\to f_{xy}$ and $\rho=1$. 
Further, in Eq.~\eqref{eq:A860} the summed
indices $m$, $n$ must satisfy $m+n\geq 1$.

If we consider the resonance-hole case, we have to insert \eqref{eq:A860}
into \eqref{eq:C007}. Then the first term on the r.h.s.
of Eq.~\eqref{eq:A860} yields $A_{xy}^S(q^2)\Upsilon^{*[0]}_+(M;\q,q_0)$ and
the second generates the contact terms, which in the present case 
can be expressed in terms of the following functions
\begin{multline}
  \label{eq:A992}
  T^{*[m,n]}_{S\pm}(\q,q_0)=
  -i\intq{p}\left\{\D^*(p+q)\right\}^{m-1}
  \left\{\D(p)\right\}^{n-1}\\
  \mp i\intq{p}\left\{\D^*(p-q)\right\}^{m-1}
  \left\{\D(p)\right\}^{n-1}~.
\end{multline}
Then, because of the vacuum subtraction, only the  $n=0$ term survives in
\eqref{eq:A860} (otherwise any dependence upon $k_F$ is lost) and of course
it must be $m\geq1$. 

The 3/2-spin resonance generates an additional complication due to
the possible presence of projection operators, as discussed in Sec.~\ref{sec:9}
(see Eq.~\eqref{eq:A920} for details): these will require the addition
of a function $B^S(q^2)$,
whose role will be clarified later. 
In conclusion, the most general 0-index LF has the structure
\begin{equation}
  \label{eq:A862}
  \begin{split}
    \Pi^{*(0)}_{xy}(M;\q,q_0)&\equiv\Pi^{*S}_{xy\pm}(M;\q,q_0)
    =A_{xy}^S(q^2)\Upsilon^{*[0]}_\pm(M;\q,q_0)
    \\
    &+B_{xy}^S(q^2)
    \Upsilon^{*[0]}_\pm(M=0;\q,q_0)
    +\sum_{m\geq1}\lambda_{m0}^S(q^2)
    T^{*[m,0]}_{S\pm}(\q,q_0)
  \end{split}
\end{equation}
with $A_{xy}^S(q^2)$,  $B_{xy}^S(q^2)$ and $\lambda_{m0}^S(q^2)$ 
to be specified according to the problem one deals with
(however $B_{xy}^S(q^2)=0$ for spin-1/2 particles).

The functions $T^{*[m,0]}_{S\pm}(\q,q_0)$, linked to the 
functions $\T$ of Eq.~\eqref{eq:A637}, are not Lorentz invariant.
Those entering our calculations are explicitly given in Appendix 
\ref{sec:AppD}.

The nucleon-hole case is more involved. Eq.~\eqref{eq:A860} still holds 
valid, provided $\D^*=\D$, but the subtraction scheme will be different, since
$\D(p+q)$ also depends upon $k_F$. Thus the contact terms will also be 
different because both the cases $m=0,n\not=0$ and $m\not=0,n=0$ 
contribute in this instance after the vacuum subtraction.

\subsection{Vector case: 1-index functions}
\label{sec:3.3}

Vector-like LFs can only arise through the combination of a scalar
and a vector vertex. Lorentz invariance would forbid such transitions,
because scalar and vectors belong to different representations of the Lorentz
group but, since the infinite nuclear medium violates covariance due to the 
presence of the $\theta$-functions, these terms
may occur. Hence in the nuclear medium a vector meson
(the $\omega$, for instance) can be converted into a $\sigma$. 

By covariance the   functions $f_{xy}^\mu$ must have the structure 
\begin{equation}
f_{xy}^\mu=\alpha t^\mu+\beta q^\mu ~,
  \label{eq:fxy}
\end{equation}
where the transverse momentum
\begin{equation}
  \label{eq:A712}
  t^\mu=p^\mu-\frac{p\cdot q}{q^2}q^\mu
\end{equation}
has been introduced ($t\cdot q=0$).
The second term on the r.h.s. of Eq.~\eqref{eq:fxy} is immediately handled, 
since the vector $q^\mu$ 
factors out of the integral and the scheme of the previous subsection 
applies, but it gives rise to a $\Pi^{*(1)\mu}_{xy}$ which is not gauge 
invariant.

Instead, the vector-like LF generated by the first term of 
$f_{xy}^\mu$ (to be called $\Pi^{*(1)\mu~\rm cons}_{xy}$)
obeys the conservation law $q_\mu\Pi^{*(1)\mu~\rm cons}_{xy}=0$
and can be cast into the form
\begin{equation}
  \label{eq:A498}
  \Pi^{*(1)\mu~\rm cons}_{xy}=\Pi^{*(1)0~\rm cons}_{xy}\N^\mu~,
\end{equation}
where we have introduced the four-component object (not a vector)
\begin{equation}
  \label{eq:A873}
  \N^\mu=\left(1,\dfrac{q_0 q^i}{\q^2}\right)~.
\end{equation}
Hence it is sufficient to compute
the 0 component of $\Pi^{*(1)\mu}_{xy}$ only.
Defining 
\begin{equation}
  \label{eq:781}
  \Q^{*V}_\pm(M;\q,q_0)=-i\intq{p}t^0
    \left\{\S^*(p+q)\S_m(p)\pm\S^*(p-q)\S_m(p)\right\}
\end{equation}
and
\begin{equation}
  \label{eq:A512}
  T^{*[m,n]}_{V\pm}(\q,q_0)=-i\intq{p}t^0\;\{\D^*(p+q)\}^{m-1}\{\D(p)\}^{n-1}
  \pm (q_0\leftrightarrow -q_0)
\end{equation}
the 0 component of the vector-like LF, using again 
\eqref{eq:A710}, takes the form 
\begin{equation}
  \label{eq:A821}
  \begin{split}
    \Pi^{*(1)0~\rm cons}_{xy\pm}&\equiv\Pi^{*V}_{xy\pm}=A^V_{xy}(q^2)\Q^{*V}_\pm
    (M;\q,q_0)+B^V_{xy}(q^2)\Q^{*V}_\pm
    (M=0;\q,q_0)\\
    &+\sum_m \lambda^V_{m0}
      T^{*[m,0]}_{V\pm}(\q,q_0)~,
      \end{split}
\end{equation}
where we have accounted for the projection operators
\eqref{eq:A920} and the parity is not specified.
Again the functions $A$, $B$ and $\lambda$ have to be specified according
to the problem.

The new function $\Q^{*V}_\pm$ is expressible in terms of the $\Upsilon^{*[n]}$
as follows
\begin{equation}
  \label{eq:A670}
  \Q^{*V}_\pm(M;\q,q_0)
  =\Upsilon^{*[1]}_\pm(M;\q,q_0)
  +\frac{1}{2}\rho q_0\Upsilon^{*[0]}_\mp(M;\q,q_0)  -
  \begin{cases}
    0\\
    \dfrac{q_0}{q^2}\T^{[0]}
  \end{cases}~,
\end{equation}
while the $T^{*[m,n]}_V$
relevant to us are listed in Appendix \ref{sec:AppD}.
In conclusion the general structure of the 1-index LF reads
\begin{equation}
  \label{eq:A327}
  \Pi^{*(1)\mu}_{xy}(M;\q,q_0)=\Pi^{*V}_{xy+}(M;\q,q_0)\N^\mu
  +q^\mu\Pi^{*S}_{xy-}(M;\q,q_0)~.
\end{equation}

Again the nucleonic case is more tricky,
because we must first replace \eqref{eq:781} with
\begin{equation}
  \label{eq:A991}
  \Q^V(\q,q_0)=-i\intq{p}t^0
    \S_m(p+q)\S_m(p)~,
\end{equation}
next express $p\cdot q$ according to \eqref{eq:A710} and finally
use directly the expression \eqref{eq:C006}.
Following exactly the same path leading to Eq.~\eqref{eq:C016} we obtain
the limiting case ($M=m$) of Eq.~\eqref{eq:A670}.

\subsection{Tensor case: 2-indices functions}
\label{sec:3.4}

The 2-indices functions, which require two vector-type vertices, can be split
into a symmetric and antisymmetric part that need to be
studied separately.

\subsubsection{Symmetric case}

\label{sec:3.4.1}

In the symmetric case, since $f^{\mu\nu}_{xy}$ is a true tensor,
Lorentz covariance imposes the structure
\begin{equation}
  \label{eq:A662}
  f^{\mu\nu}_{xy~{\rm symm}}=a_1 \left(q^2g^{\mu\nu}
    -q^\mu q^\nu\right)+a_2 t^\mu t^\nu+a_3 (t^\mu q^\nu+
  q^\mu t^\nu)+a_4 q^\mu q^\nu
\end{equation}
the $a_i$ being Lorentz invariants.

The first two terms of \eqref{eq:A662} correspond to a conserved current.
Let us denote the associated LF by $\Pi^{*(2)\mu\nu
~\rm cons}_{xy~\rm symm}(M;\q,q_0)$, which
can be split into a longitudinal ($\Pi^{*L}_{xy}$) and a 
transverse ($\Pi^{*T}_{xy}$)
polarisation propagators, defined according to
\begin{subeqnarray}
  \label{eq:A880}
  \Pi^{*L}(M;\q,q_0)&=&\Pi^{*(2)00~\rm cons}_{xy~\rm symm}(M;\q,q_0)~,\\
  \Pi^{*T}(M;\q,q_0)&=&\left(\delta_{ij}-\frac{q_iq_j}{\q^2}\right)
    \Pi^{*(2)ij ~\rm cons}_{xy~\rm symm}(M;\q,q_0)
\end{subeqnarray}
and, in a compact notation,
\begin{multline}
  \label{eq:A881}
  \Pi^{*(2)\mu\nu~\rm cons}_{xy~\rm symm}(M;\q,q_0)=\\
  \left(
  \begin{tabular}{c|c}
    $\Pi^{*L}$&$\frac{q_0q_i}{\q^2}\Pi^{*L}$\\
    \hline
    $\frac{q_0q_j}{\q^2}\Pi^{*L}$
    &$\frac{q_0^2}{\q^2}\Pi^{*L}\frac{q_iq_j}{\q^2}
    +\frac{1}{2}\Pi^{*T}\left(\delta_{ij}-\frac{q_iq_j}{\q^2} \right)
    $
  \end{tabular}\right)~.
\end{multline}

The first term in the r.h.s. of \eqref{eq:A662} is easily handled, 
since it reduces to the 0-index case. 
The second instead requires the introduction of two new
quantities with the associated contact terms. We thus define
\begin{align}
  \Q^{*L}_\pm&(M;\q,q_0)=\intt{p}\frac{(t^0)^2}{2E_p}
  \frac{\theta(k_F-\p)}{(p+q)^2-M^2+i\eta}\Biggm|_{p_0=E_p}\pm
  ~(q\longleftrightarrow -q)\nonumber\\
  &=\Upsilon^{*[2]}_\pm(M;\q,q_0)+
  q_0\rho \Upsilon^{*[1]}_\mp(M;\q,q_0)\nonumber\\
  &+\frac{1}{4}q_0^2\rho^2\Upsilon^{*[0]}_\pm(M;\q,q_0)
  -\begin{cases}
    \dfrac{q_0^2\rho}{2q^2}\T^{[0]}\\
    \dfrac{q_0}{q^2}\left(1-\frac{\q^2}{q^2}\right)\T^{[1]}
  \end{cases}
  \label{eq:A121}\\
\intertext{and}
    \Q^{*T}_\pm&(M;\q,q_0)\nonumber\\
    &=\frac{1}{\q^2}\intt{p}
    \frac{\p^2\q^2-({\bf p}\cdot{\bf q})^2}{2E_p}
    \frac{\theta(k_F-\p)}{(p+q)^2-M^2+i\eta}\Biggm|_{p_0=E_p\nonumber}\\
    &\pm~(q\longleftrightarrow -q)\nonumber\\
    &=-\frac{q^2}{\q^2}\Biggl\{\Upsilon^{*[2]}_\pm(M;\q,q_0)
    +q_0\rho \Upsilon^{*[1]}_\mp(M;\q,q_0)\nonumber\\
    &+\left(\frac{1}{4}q^2\rho^2+\frac{m^2\q^2}{q^2}\right)
    \Upsilon^{*[0]}_\pm(M;\q,q_0)\Biggr\}
    +\begin{cases}
      \dfrac{q^2\rho}{2\q^2}\T^{[0]}\\~\\
      \dfrac{q_0}{q^2}
      {\mathfrak T}^{[1]}
    \end{cases}
  \label{eq:A215}
\end{align}
together with the contact terms
\begin{align}
\label{eq:A711}
  T^{*[m,n]}_{L\pm}(\q,q_0)&=-i\intq{p}(t^0)^2\;\{\D^*(p+q)\}^{m-1}
  \{\D(p)\}^{n-1}
  \pm (q_0\leftrightarrow -q_0)\\
  \intertext{and}
  T^{*[m,n]}_{T\pm}(\q,q_0)&=\frac{-i}{\q^2}\intq{p}
  \left[\p^2\q^2-({\bf p}\cdot{\bf q})^2\right]\times\\
  &\{\D^*(p+q)\}^{m-1}\{\D(p)\}^{n-1}
  \nonumber
  \pm (q_0\leftrightarrow -q_0)~.
\end{align}
Thus, applying \eqref{eq:A710} and \eqref{eq:A862}, we obtain
\begin{align}
  \label{eq:A878}
  \Pi^{*L}_{xy}(M;\q,q_0)&=-\q^2\Pi^{*S}_{xy}(M;\q,q_0)
  +A^{LT}_{xy}(q^2)\Q^{*L}_+(M;\q,q_0)\\
  &\nonumber
  +B^{LT}_{xy}(q^2)\Q^{*L}_+(M=0;\q,q_0)+\sum_m \lambda_{m0}^{LT}(q^2)
  T^{*[m,0]}_{L+}(\q,q_0)\\
  \label{eq:A567}
  \Pi^{*T}_{xy}(M;\q,q_0)&=-2q^2\Pi^{*S}_{xy}(M;\q,q_0)
  +A^{LT}_{xy}(q^2)\Q^{*T}_+(M;\q,q_0)\\
  &\nonumber
  +B^{LT}_{xy}(q^2)\Q^{*T}_+(M=0;\q,q_0)+\sum_m \lambda_{m0}^{LT}(q^2)
  T^{*[m,0]}_{T+}(\q,q_0)
\end{align}
(note that the same coefficients $A^{LT}$, $B^{LT}$ and $\lambda^{LT}$ enter
in both $\Pi^{*L}$ and $\Pi^{*T}$).
The above relations give the structure of the longitudinal and transverse
LFs and thus fully describe
$\Pi^{*(2)\mu\nu
~\rm cons}_{xy~\rm symm}$ through \eqref{eq:A881}. Finally, the remaining terms
of \eqref{eq:A662} can be 
reduced to simpler cases and one gets the final result
\begin{equation}
  \label{eq:A663}
  \begin{split}
    \Pi^{*(2)\mu\nu}_{xy~\rm symm}(M;\q,q_0)&=\Pi^{*(2)\mu\nu
      ~\rm cons}_{xy~\rm symm}(M;\q,q_0)\\&+(q^\mu\N^\nu+q^\nu\N^\mu)
    \Pi^{*V}_{xy-}(M;\q,q_0)\\&+q^\mu q^\nu \tilde\Pi^{*S}_{xy+}(M;\q,q_0)~.
  \end{split}
\end{equation}
The last term on the r.h.s. of \eqref{eq:A663}, 
namely $\tilde\Pi^{*S}_{xy+}$, corresponds to a LF with the
same structure of $\Pi^{*S}_{xy+}$ but with different ingredients:
these will be called $\tilde A^S$, $\tilde B^S$ and $\tilde \lambda^S_{m0}$,
respectively, to avoid confusion.

Again the nucleon-hole case must be handled separately, 
by using the analogous
of Eq.~\eqref{eq:A991}. Clearly now, 
at variance of  the resonance case, only the
$\Q^{*L,T}_+$ terms exist, and a straightforward  calculation
shows that Eqs.~\eqref{eq:A121} and \eqref{eq:A215} still hold valid.

\subsubsection{Antisymmetric case}

An antisymmetric tensor should have the form
\begin{equation}
  \label{eq:A667}
  f^{*\mu\nu}_{xy~\rm antisymm}=b_1 (t^\mu q^\nu-
  q^\mu t^\nu)+b_2 \epsilon^{\mu\nu\lambda\rho}p_\lambda q_\rho
\end{equation}
and the general structure of $\Pi^{*(2)\mu\nu}_{xy~\rm antisymm}$ will 
accordingly be
\begin{multline}
  \label{eq:A685}
  \Pi^{*(2)\mu\nu}_{xy~\rm antisymm}(M;\q,q_0)=
  (q^\mu\N^\nu-q^\nu\N^\mu)\Pi^{*V}_{xy-}(M;\q,q_0)\\
  +q_\lambda \N^\sigma g_{\sigma\rho} 
  \epsilon^{\mu\nu\rho\lambda}\tilde\Pi^{*V}_{xy-}(M;\q,q_0)~.
\end{multline}
Again the function $\tilde\Pi^{*V}_{xy-}$ entering the above has
the same structure given by Eq.~\eqref{eq:A821}, but with the functions
$A^V$, $B^V$ and $\lambda^V_{m0}$ replaced by 
$\tilde A^V$, $\tilde B^V$ and $\tilde\lambda^V_{m0}$.

\section{Analytic evaluation of $U^{*[0]}_{rel}(M;\q,q_0)$}

\label{sec:4}

In this section we explicitly compute the function $U^{*[0]}_{rel}$
defined by Eq.~\eqref{eq:C009}
(the other two functions $U^{*[1]}_{rel}$ and $U^{*[2]}_{rel}$
can be obtained along the same path and the results are reported  
in Appendix \ref{sec:appA}). 

Assuming here $\Im~ q_0\not=0$  we get
\begin{equation}
  \begin{split}
    \label{eq:AC103}
    U^{*[0]}_{rel}&=\intt{p}\frac{1}{2E_p}
    \frac{\theta(k_F-\p)}{2E_pq_0-2{\bf p}\cdot{\bf q}+q^2\rho}\\
    &=\frac{1}{16\pi^2\q}\int\limits_0^{k_F}\frac{p\,dp}{E_p}
    \log\frac{q^2\rho +2 p\q +2q_0E_p}{q^2\rho-2 p\q+2q_0E_p}~.
  \end{split}
\end{equation}
Note that the dependence upon $M$ is  fully embodied in the inelasticity
parameter $\rho$.
Integration by parts yields
\begin{equation}
  \label{eq:AC012}
  \begin{split}
    U^{*[0]}_{rel}&=\frac{1}{16\pi^2\q}E_F\log\frac{q^2\rho+2k_F\q+2E_Fq_0}
    {q^2\rho-2k_F\q+2E_Fq_0}\\&-\frac{1}{4\pi^2}\int\limits_0^{k_F}
    dp\;\frac{q^2\rho E_p+2m^2q_0}
    {\left(q^2\rho-2p\q+2E_pq_0\right)
    \left(q^2\rho+2p\q+2E_pq_0\right)}~.
  \end{split}
\end{equation}

The integrand in \eqref{eq:AC012} displays four poles, located at 
$p=\pm y^*_\pm$.
Defining 
\begin{equation}
  \label{eq:A522}
  Q_\pm=(M\pm m)^2-q^2
\end{equation}
and 
\begin{equation}
  \label{eq:AC405}
  \Delta^*=\sqrt{\rho^2-\frac{4m^2}{q^2}}=-\frac{\sqrt{Q_+Q_-}}{q^2}
  =-\frac{\sqrt{\left[(M-m)^2-q^2\right]
      \left[(M+m)^2-q^2\right]}}{q^2}
\end{equation}
(with the chosen sign $\Delta^*$ is positive in the space-like region)
we obtain for the poles the expression
\begin{equation}
  \label{eq:AC006}
  y_\pm^*=\frac{\q}{2}\rho
  \pm \frac{q_0}{2}\Delta^*~.
\end{equation}
The  $\pm y^*_\pm$ are the branch points defining
the region where,
for real $q_0$, \eqref{eq:AC012} develops an imaginary part. 
In particular the lowest positive
branch point is just the lowest possible longitudinal momentum for
the occurrence of a resonance-hole pair and consequently coincides
(up to a sign) with the $y$ scaling variable. Indeed
$-y^*_-$, taken at $M=m$, is just the $y$ scaling variable for the 
relativistic Fermi gas~\cite{CeDoMo-97}. 

The integrand of \eqref{eq:AC012} can be split according to
\begin{multline}
  \frac{E_p}{32\pi^2\q}
  \left\{\frac{1}{p+y^*_-}+\frac{1}{p+y^*_+}-\frac{1}{p-y^*_-}
    -\frac{1}{p-y^*_+}
      \right\}\\
      +\frac{1}{32\pi^2\q}
      \left\{\frac{R^*_-}{p+y^*_-}-\frac{R^*_+}{p+y^*_+}-\frac{R^*_-}{p-y^*_-}
    +\frac{R^*_+}{p-y^*_+}\right\}~,
\end{multline}
where
\begin{equation}
  \label{eq:AC008}
  R_\pm^*=\frac{\q}{2}\Delta^*\pm\frac{q_0}{2}\rho
\end{equation}
has the property
\begin{equation}
  \label{eq:A401}
  (R_\pm^*)^2=m^2+(y_\pm^*)^2
\end{equation}
and is trivially linked to the scaling variable $\psi^*$ used in 
Refs.~\cite{Am-al-99,AlBaDoMo-01,AlBaDoMo-03,BaCaDoMa-04}
by
\begin{equation}
  \label{eq:C999}
  R^*_-=m+(E_F-m){\psi^*}^2~.
\end{equation}
We can now easily compute explicitly \eqref{eq:AC012}, getting
\begin{equation}
  \label{eq:AC004}
  \begin{split}
    &U^{*[0]}_{rel}(M;\q,q_0)=-\frac{\rho}{(4\pi)^2}\log\frac{k_F+E_F}{m}\\
    &-\frac{E_F}{16\pi^2\q}\log\frac{q^2\rho-2k_F\q+2E_Fq_0}
    {q^2\rho+2k_F\q+2E_Fq_0}\\
    &+\frac{q_0\rho}{64\pi^2\q} \log\frac{(k_F-y_-^*)^2
      (m^2+k_Fy_+^*+E_FR^*_+)(m^2-k_Fy_+^*-E_FR^*_+)}{(k_F+y_+^*)^2
      (m^2+k_Fy_-^*+E_FR^*_-)(m^2-k_Fy_-^*-E_FR^*_-)}\\
    &-\frac{\Delta^*}{64\pi^2}
    \log\frac{m^4(k_F-y_-^*)^2(k_F+y_+^*)^2}
    {\left[m^4-(k_F y^*_-+E_FR^*_-)^2\right]
      \left[m^4-(k_F y^*_++E_FR^*_+)^2\right]}
  \end{split}
\end{equation}
or, with some algebra, 
\begin{equation}
  \label{eq:A554}
  \begin{split}
    &U^{*[0]}_{rel}(M;\q,q_0)=-\frac{y^*_-+y^*_+}{16\pi^2\q}
    \log\frac{k_F+E_F}{m}+\frac{R^*_+-R^*_-+2E_F}{64\pi^2\q}\times\\
    & \times\log\frac{(k_F-y_-^*)^2
      (m^2+k_Fy_+^*+E_FR^*_+)(m^2-k_Fy_+^*-E_FR^*_+)}{(k_F+y_+^*)^2
      (m^2+k_Fy_-^*+E_FR^*_-)(m^2-k_Fy_+^*-E_FR^*_+)}\\
    &-\frac{R^*_++R^*_-}{64\pi^2\q}
    \log\frac{m^4(k_F-y_-^*)^2(k_F+y_+^*)^2}
    {\left[m^4-(k_F y^*_-+E_FR^*_-)^2\right]
      \left[m^4-(k_F y^*_++E_FR^*_+)^2\right]}~,
  \end{split}
\end{equation}
which depends, but for the overall factor
$\q^{-1}$, only upon the two scaling variables $y^*_\pm$. 

Finally, with $E^*_p=\sqrt{M^2+p^2}$
(hence $E^*_{k_F\pm q}=\sqrt{M^2+(k_F\pm q)^2}$) and bringing back $q_0$ to
the real axis, Eq.~\eqref{eq:AC004} reduces to the compact form
\begin{equation}
  \label{eq:AC209}
  U^{*[0]}_{rel}(M;\q,q_0)=-\frac{\rho}{16\pi^2}l_1
  +\frac{2E_F+\rho q_0}{32\pi^2\q}l_2
  -\frac{\Delta^*}{64\pi^2}l_3~,
\end{equation}
having defined the logarithmic functions 
\begin{subeqnarray}
  \label{eq:A300}
  l_1&=&\log\frac{k_F+E_F}{m}\\
  l_2&=&\log
      \left|\frac{
      q_0+E^*_{k_F-q}+E_F}{
      q_0+E^*_{k_F+q}+E_F}\right|+\log
      \left|\frac{q_0-E^*_{k_F-q}+E_F
      }{q_0-E^*_{k_F+q}+E_F
      }\right|+i\pi k_2\\
  l_3&=&\log\left|\frac{(2k_F+q_0\Delta^*)^2-\q^2\rho^2}
      {(2k_F-q_0\Delta^*)^2-\q^2\rho^2}\right|+\log\left|
      \frac{(\rho k_F-E_F\Delta^*)^2-\dfrac{4\q^2m^4}{q^4}}
      {(\rho k_F+E_F\Delta^*)^2-\dfrac{4\q^2m^4}{q^4}}\right|
    \nonumber
    \\&+&i\pi k_3   
\end{subeqnarray}
if $\Delta^*$ is real and
\begin{align}
  \label{eq:A122}
    l_3&=2i\arctan\frac{4k_Fq_0|\Delta^*|}{d_1}
    -2i\arctan\frac{2k_FE_F\rho|\Delta^*|}{d_2}\\
    &-2i\pi\theta(d_1)\biggl\{\theta(q_0)+
    \theta(-q_0)\theta(M^2+m^2-2k_F^2+\q^2)\nonumber\\
    &\qquad\qquad\times{\rm sgn}(q_0+\sqrt{M^2+m^2-2k_F^2+\q^2})\biggr\}
    \nonumber\\
    &+2i\pi\theta\left(d_2\right)\left\{\theta(q_0)+{\rm sgn}
    \left(q_0-\sqrt{M^2+\q^2+\frac{m^4}{m^2+2k_F^2}}\right)\right\}\nonumber\\
    d_1&=4k_F^2-q_0^2|\Delta^*|^2-\q^2\rho^2\\
    d_2&=\rho^2k_F^2+E_F^2|\Delta^*|^2
      -\frac{4m^4\q^2}{q^4}
\end{align}
if  $\Delta^*=i|\Delta^*|$ is purely imaginary 
(i.e., for $(Q^T_-)^2<q_0^2<(Q^T_+)^2$, see Eq.~\eqref{eq:AC022}).
In this case the function $U^{*[0]}_{rel}$ is 
manifestly real. The two $\theta$ functions force $l_3$ to be continuous.
In Eqs.~\eqref{eq:A300} 
$k_2$ and $k_3$ are integer that can be fixed by analytic extension
or by checking the integral in some suitable points. They  
will be specified in the next section and found to
depend only upon the analytic structure of the
logarithms and thus remain the same for the whole set of functions
$ U^{*[n]}_{rel}$.

\section{The response region}
\label{sec:5}

As previously mentioned, the singularities of  \eqref{eq:AC004}, independent
of $f_{xy}$, fully determine the response 
regions for particle(resonance)-hole(antiparticle) excitations. 
Further, the values of the $k_i$ in \eqref{eq:A300}
are also independent of $f_{xy}$ and will be determined in this section,
where we consider a real $q_0$ up to a vanishingly small imaginary part 
$\pm i\eta$.

To set the response regions 
first consider the branch points associated with the excitation of a 
resonance in the free space. They are located at
\begin{equation}
  \label{eq:AC022}
  Q^T_\pm=\sqrt{\q^2+(M\pm m)^2}~,
\end{equation}
which define the boundaries of the regions
where the production of an anti\-resonance-particle plus the emission 
($q_0<-Q^T_+$) or the absorption ($q_0>Q^T_+$) of a probe
or the excitation of a particle to a resonance
($-Q^T_-<q_0<Q^T_-$) are allowed.
Accordingly $Q^T_+$ and $Q^T_-$ are usually referred to as the threshold and 
pseudo-threshold, respectively.
These singularities clearly stem from the presence of the Dirac sea.

Next we discuss the domain where  the response of the system to a probe
accounts for the effect of the medium,  hence the effect of the Fermi sea.
This is fixed by the logarithmic singularities of  $U^*_{rel}$, which turn out
to be located at
\begin{equation}
  \label{eq:AC021}
  Q^B_{\pm,\pm}(p)=\pm E^*_{p\pm q}-E_p~,
\end{equation}
that, for $p=k_F$, fix the boundaries (hence the label $B$)
of the resonance-hole and antiresonance-hole regions. 
An easy check yields the ordering
\begin{equation}
  \label{eq:AC044}
  (Q^B_{-+})^2>(Q^B_{--})^2\geq(Q^T_+)^2
  >(Q^T_-)^2\geq(Q^B_{++})^2>(Q^B_{+-})^2
\end{equation}
(uniformly in $p$). 
Also, it is found that
\begin{equation}
  \label{eq:AC098}
  Q^B_{+\pm}(p)>-\q~,\qquad\qquad Q^B_{-\pm}(p)<-\q
  \qquad\forall p
\end{equation}
and
\begin{equation}
  \label{eq:AC099}
  (Q^T_{\pm})^2>\q^2~.
\end{equation}

Note that six critical values of $\q$ exist, corresponding to the
various intersections of the boundaries. We have
\begin{subeqnarray}
  \label{eq:AC047}
  (Q^B_{--})^2=(-Q^T_+)^2&\qquad\text{for}\qquad&
  q=q_{\rm cr}^{(1)}=\frac{M+m}{m}k_F~,\\
  (Q^B_{++})^2=(Q^T_-)^2&\qquad\text{for}\qquad&
  q=q_{\rm cr}^{(2)}=\frac{M-m}{m}k_F~.
\end{subeqnarray}
Concerning the sign of the singularities one finds that
$Q^B_{-\pm}<0$ and $Q^B_{++}>0$ for any value of $k_F$. 
Instead, one has
$Q^B_{+-}>0$ if $k_F<k_F^{\rm cr}$ with 
\begin{equation}
  \label{eq:AC048}
  k_F^{\rm cr}=\sqrt{M^2-m^2}~,
\end{equation}
while, when $k_F>k_F^{\rm cr}$, $Q^B_{-+}$ is negative in the interval
$q_{\rm cr}^{(3)}<q<q_{\rm cr}^{(4)}$, being
\begin{subeqnarray}
  \label{eq:AC049}
  q_{\rm cr}^{(3)}
  =&k_F-\sqrt{m^2-M^2+k_F^2}&=k_F-\sqrt{k_F^2-(k_F^{\rm cr})^2}~,\\
  q_{\rm cr}^{(4)}
  =&k_F+\sqrt{m^2-M^2+k_F^2}&=k_F+\sqrt{k_F^2-(k_F^{\rm cr})^2}~.
\end{subeqnarray}
This occurrence deserves a comment: in fact, in the limiting case $M=m$,
$k_F^{\rm cr}=0$ and we are always in the second case, with
$q_{\rm cr}^{(3)}=0$ and $q_{\rm cr}^{(4)}=2k_F$. In this region 
the response function at fixed $\q$ and as a function of $q_0$ 
has a discontinuity in the derivative.

Finally other critical points arise in connection with the light cone.
It is obvious that $(Q^T_\pm)^2>\q^2$, hence we must only inquire 
about $Q^B_{+\pm}$. We find
\begin{eqnarray}
  Q^B_{+-}
  \begin{cases}
    >\q&\text{for $q<q_{\rm cr}^{(5)}$}\\
    <\q&\text{for $q>q_{\rm cr}^{(5)}$}
  \end{cases}
  \label{eq:AC050}
  \\
  Q^B_{++}
  \begin{cases}
    >\q&\text{for $q<q_{\rm cr}^{(6)}$}\\
    <\q&\text{for $q>q_{\rm cr}^{(6)}$}~,
  \end{cases}
\end{eqnarray}
where two new critical points appear, namely
\begin{eqnarray}
  \label{eq:AC051}
  q_{\rm cr}^{(5)}&=&\frac{(M^2-m^2)(E_F-k_F)}{2m^2}~,\\
  q_{\rm cr}^{(6)}&=&\frac{(M^2-m^2)(E_F+k_F)}{2m^2}~.
\end{eqnarray}

Now we  discuss the response regions in the $(q+0,|\q|)$ plane
and fix the corresponding values of
$k_2$ and $k_3$ appearing in \eqref{eq:A300}. 

Consider first the negative time-like region. Here for $q_0<-Q^T_+$ 
the antiresonance-particle production is allowed in the vacuum,
but is ruled out by the renormalisation, which subtracts out 
the vacuum effects. 
It is, in any case, Pauli-blocked by the Fermi 
sea in the region spanned by $q_0=-\sqrt{({\bf p}+{\bf q})^2+M^2}-E_p$ 
for $\p<k_F$, namely
$$\max\{Q^B_{--}(p)\}>q_0>\min\{Q^B_{-+}(p)\}
~.$$ 
The boundaries of the permitted  response region are then found to be
\begin{equation}
  \label{eq:A201}
  Q^B_{-+}(k_F)<q_0<
  \begin{cases}
    Q^T_+\qquad&\text{for~}q<q_{\rm cr}^{(1)}\\
    Q^B_{--}(k_F)&\text{for~}q>q_{\rm cr}^{(1)}
    \end{cases}~~.
\end{equation}

The resonance-hole region instead corresponds to $q_0>-\q$
and lives partly in the space-like region and partly in the time-like one. 
The allowed values of $q_0$ span the interval
$$\min\{Q^B_{+-}(p)\}<q_0<\max\{Q^B_{++}(p)\}~,$$
again with $\p<k_F$, the associated boundaries being
\begin{equation}
  \label{eq:A311}
  Q^B_{+-}(k_F)<q_0<
  \begin{cases}
    Q^T_-\qquad&\text{for~}q<q_{\rm cr}^{(2)}\\
    Q^B_{++}(k_F)&\text{for~}q>q_{\rm cr}^{(2)}~.
    \end{cases}
\end{equation}

In order to fix $k_2$ and $k_3$, which are integer and constant 
inside all the response regions, it is then sufficient to
evaluate the imaginary part of $U^{*}_{rel}$
in some particularly simple point of the response regions. 

Consider first the regions $Q^B_{++}<q_0<Q^T_-$ and $Q^B_{-+}<q_0<-Q^T_+$.
Here a convenient point is $\q=0$, where
$$\Im~U^{*[0]}_{rel}=\mp \pi\intt{p}\frac{1}{2E_p}
\theta(k_F-p)\delta(2E_p q_0+\rho q_0^2)\Biggm|_{|{\mathbf q}|=0}~,
$$
which is non-vanishing only in the regions quoted above,
its value being
$$\Im~U^{*[0]}_{rel}=\mp \frac{|\Delta^*|}{16\pi}\Biggm|_{|{\mathbf q}|=0}~.$$
This result deserves a few comments. First, from the definition it follows
that the imaginary part of $U^{*[0]}_{rel}$ must have the sign
$\mp$ according to whether $q_0\to q_0\pm i\eta$, in accord with the 
above outcome.
Then, since $\Delta^*$ is negative in the time-like region, we obtain
\begin{subeqnarray}
  \label{eq:A592}
  k_2&=&0\\
  k_3&=&\mp 4
\end{subeqnarray}
for $Q^B_{++}<q_0<Q^T_-$ or $Q^B_{-+}<q_0<-Q^T_+$.

Next we consider the region $Q^B_{+-}<q_0<Q^B_{++}$. Here,
choosing a very small value for $k_F$,
the point $q_0=\sqrt{M^2+\q^2}-m$ surely lies in the desired
region, and one gets
$$\Im~U^{*[0]}_{rel}=\mp \frac{1}{4\pi^2}\int\limits_0^{k_F}\frac{p^2 dp}{2E_p}
\int\limits_{-1}^1 dx \delta(2(E^*_q-m)(E_p-m)-2 p q x)~.
$$
Being $k_F$ (and hence $p$) small,
the argument of the $\delta$-function vanishes at
$$\bar x=\frac{p}{2\q}\left(\frac{E^*_q}{m}-1\right)$$
which is less than one. Thus the angular integration becomes trivial and 
we get
$$\Im~U^{*}[0]_{rel}=\mp\frac{E_F-m}{16\pi\q}~.$$
Since this result should be a combination of the coefficients 
of $l_2$ and $l_3$ then
\begin{subeqnarray}
  \label{eq:A595}
  k_2&=&\mp 1\\
  k_3&=&\mp 2~.
\end{subeqnarray}

Finally we consider the region $Q^B_{-+}<q_0<Q^B_{--}$. Here we choose
$q_0=-E^*_q-m$ and, following the same steps as before, under the 
same assumptions, we find that Eqs.~\eqref{eq:A595} hold again.

\begin{figure}[htbd]
  \begin{center}
    \leavevmode
    \hbox{
      \epsfig{file=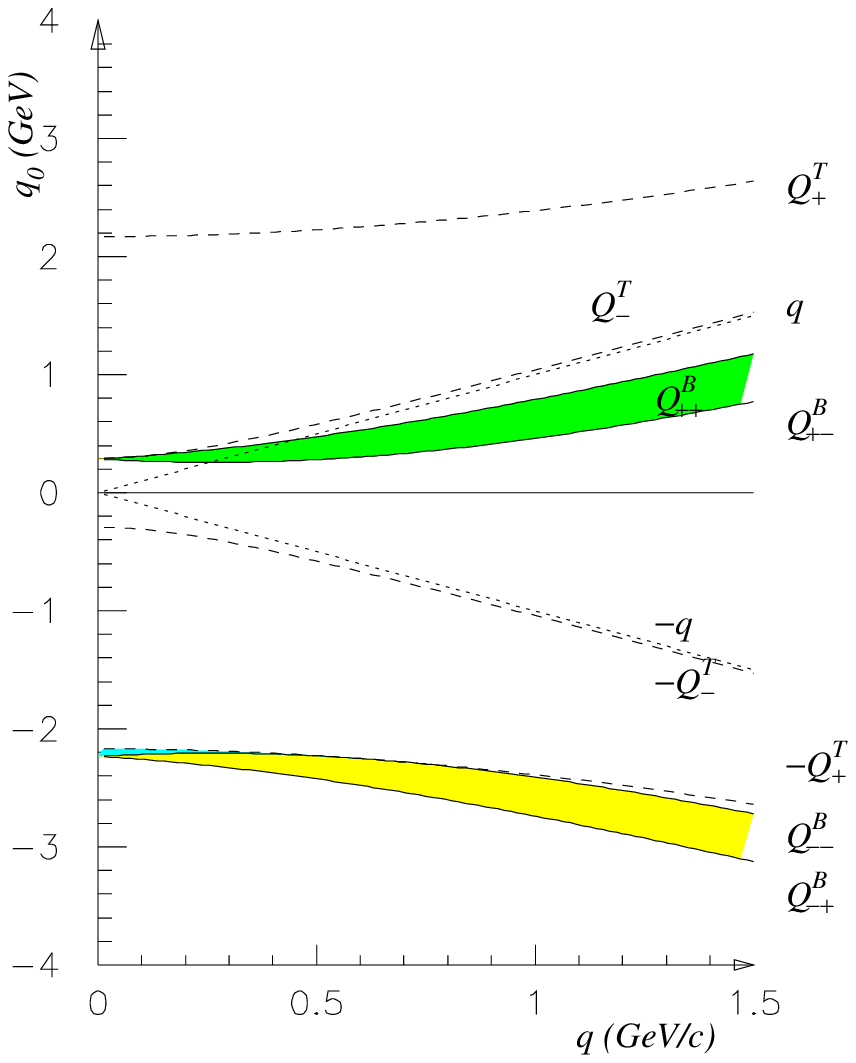,height=14cm,width=12cm}
      }
    \caption{The singularities $Q^B$ and $Q^T$ of the function 
      $U^{*[0]}_{rel}$ 
      in the $(\q,q_0)$ plane. Here, having in mind nucleons and $\Delta$'s, 
      we have $m=0.938$ GeV and $M=1.232$ GeV. Solid lines denote 
      the four singularities $Q^B_{\pm \pm}$ and dashed lines 
      the $\pm Q^T_\pm$. Dotted lines denote the light cone. 
      $k_F=0.264$ GeV/c. 
Light grey (yellow if in colors) region: $Q^B_{--}>q_0>Q^B_{-+}$;
medium grey (cyan) region: $-Q^T_->q_0>Q^B_{--}$;
dark grey (green) region: $Q^B_{++}>q_0>Q^B_{+-}$; 
very dark (red) region (not visible in this figure): $Q^T_->q_0>Q^B_{++}$.
      }
    \label{fig:1}
  \end{center}
\end{figure}

The singularities, and the response regions,
in the plane $\q,q_0$ are displayed in Fig.~\ref{fig:1}
(left and right panels) for
 two different values of $k_F$, one below (left) and one well
above (right) the 
critical value of $k_F$.

\begin{figure}[htbd]
  \begin{center}
    \leavevmode
    \hbox{
      \epsfig{file=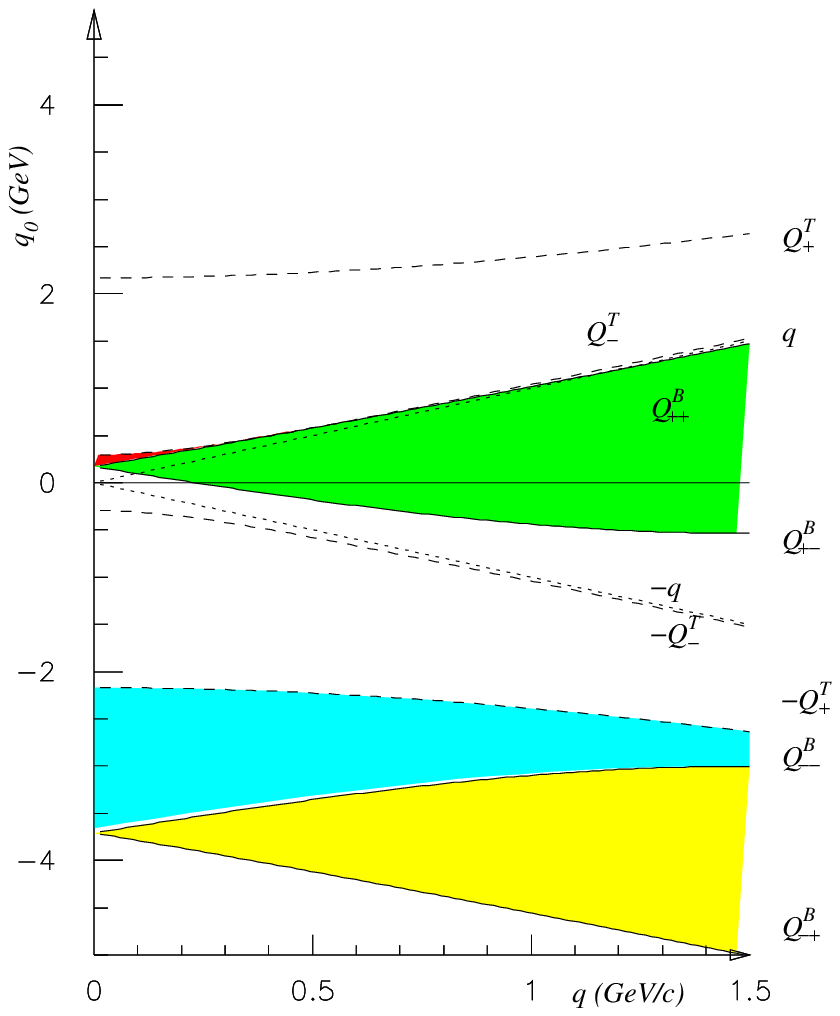,height=14cm,width=12cm}
      }
    \caption{As fig. \protect\ref{fig:1} but with $k_F=1.5$ GeV/c, in order
      to emphasize the region $Q^T_->q_0>Q^B_{++}$, otherwise invisible.}
    \label{fig:2}
  \end{center}
\end{figure}

\section{The limiting cases}
\label{sec:6}

Having determined the response regions and the integers
$k_2$ and $k_3$ we are now able to derive the response
functions in the most general case.
In this section we consider some specific limiting cases.

\subsection{The case $M=m$}
\label{sec:6a}

Here
\begin{eqnarray}
  \label{eq:A272}
  Q^T_-&=&\q~,\\
  Q^T_+&=&\sqrt{4m^2+\q^2}
\end{eqnarray}
and the critical points $q_{\rm cr}^{(1)}$ and $q_{\rm cr}^{(2)}$
occur at $2k_F$ and $0$ respectively. Furthermore, $k_F^{\rm cr}=0$
and thus  $Q^B_{+-}(k_F)=E_{k_F-q}-E_F$ is always negative
between $q_{\rm cr}^3=0$ and $q_{\rm cr}^4=2k_F$.
Finally the $Q^B_{+\pm}(k_F)$ always live in the space-like region.

The expressions for $U^{*}_{rel}$ given in Sec.~\ref{sec:4}
are still valid, provided we set $M=m$, which implies
$  \Delta\equiv \Delta^*=\sqrt{\frac{q^2-4m^2}{q^2}}$
and
$  \rho=1$.
Concerning the imaginary parts, nothing changes in the particle-antiparticle
domain, while in the particle-hole region, since $q_{\rm cr}^2$
vanishes, the response is confined to the range
$$Q^B_{+-}(k_F)<q_0<Q^B_{++}(k_F)$$
and $k_2$ and $k_3$ are given by \eqref{eq:A595}.

\subsection{The case $m=0$}
\label{sec:6b}

This case may correspond to the excitation of a light quark
to an $s$ or $c$ quark in a Quark Gluon Plasma.

Here
\begin{equation}
  \label{eq:A275}
  Q^T_{\pm}\equiv E^*_q~,
\end{equation}
while $q_{\rm cr}^{(1)}$ and $q_{\rm cr}^{(2)}$ tend to infinity: accordingly
$Q^T_{\pm}$ and $Q^B_{+-}$ never coincide.
Furthermore $k_F^{\rm cr}=M$. Finally
it is immediately seen that $q_{\rm cr}^{(6)}\to\infty$, 
implying $Q^B_{++}>\q~\forall k_F,~M$, while
$q_{\rm cr}^{(5)}=M^2/(4k_F)$. Thus the response region is represented by the 
intervals 
\begin{eqnarray*}
  Q^B_{+-}(k_F)&<q_0<&Q^B_{++}(k_F)\\
  Q^B_{-+}(k_F)&<q_0<&Q^B_{--}(k_F)~,
\end{eqnarray*}
where Eq.~\eqref{eq:A595} holds, and by
\begin{eqnarray*}
  Q^B_{++}(k_F)&<q_0<&Q^T_-\\
  Q^B_{--}(k_F)&<q_0<&-Q^T_+~,
\end{eqnarray*}
where instead \eqref{eq:A592} is valid.

Concerning the LF, since now 
$\rho=\Delta^*= 1-\frac{M^2}{q^2}$
and
$  y^*_\pm=\frac{1}{2}\left(1-\frac{M^2}{q^2}\right)(\q\pm q_0)$,
we end up with the expression
\begin{equation}
  \label{eq:A280}
  \begin{split}
    U^{*[0]}_{rel}\bigm|_{m=0}&=\frac{(2k_F+q_0)q^2-M^2q_0}{32\pi^2\q q^2}
    \left\{\log\left|\frac{M^2-2k_F(q_0+\q)-q^2}{M^2-2k_F(q_0-\q)-q^2}
      \right|+i\pi k_2\right\}\\
    &-\frac{q^2-M^2}{32\pi^2 q^2}\left\{\log\left|\frac{(M^2-2k_Fq_0-q^2)^2
        -4k_F^2\q^2}{(M^2-q^2)^2}\right|+i\pi k_3\right\}~.
  \end{split}
\end{equation}

\subsection{The case $M=m=0$.}
\label{sec:6c}

Finally we consider the extreme situation where both masses vanish.
Here the values $Q^T_\pm$ coincide with the light cone, while
$Q^B_{\pm\pm}=\pm |k_F\pm \q|-k_F$
and inside the response region only the case of Eq.~\eqref{eq:A595} occurs.

The expression of $U^{*}_{rel}$ further simplifies to
\begin{equation}
  \label{eq:A282}
    \begin{split}
    U^{*[0]}_{rel}\bigm|_{m=0}&=\frac{2k_F+q_0}{32\pi^2\q }
    \left\{\log\left|\frac{q^2+k_F(q_0+\q)}{q^2+2k_F(q_0-\q)}
      \right|+i\pi k_2\right\}\\
    &-\frac{1}{32\pi^2}\left\{\log\left|\frac{\q^2-(2k_F+q_0)^2}{q^2}\right|
      +i\pi k_3\right\}~.
  \end{split}
\end{equation}

\section{The spin 1/2 resonances}
\label{sec:7}

In this Section we explore the LFs associated to the excitations
of the nucleon, addressing first the simpler case of the spin 1/2 resonances 
(e.g. the Roper ($N^*1440$) resonance). 
For sake of simplicity we disregard isospin, which simply 
yields a numerical factor.

\subsection{The 0-index functions}
\label{sec:7.1}

Here the vertices we consider, beyond the identity,
carry some $\gamma$-matrix structure of the type $\Not p$ or $\Not q$
times, eventually, a $\gamma_5$. Owing to the mass shell condition for
the nucleon, $\Not p=(\Not p-m)+m$ can be replaced by $m$, since 
$\Not p-m$ cancels with the nucleon propagator, leaving a
$k_F$-independent term subtracted out by the renormalisation.
The nucleon-hole case requires a separate discussion.

Similarly,
$\Not q=(\Not p+\Not q-M)-(\Not p-m)-(M-m)$ leaves us
with the identity times $M-m$ plus, however, a contact term, an
occurrence reflecting our ignorance about the off-shell reaction
mechanisms. Actually the $\Not q$ vertex is redundant 
as far as  
the imaginary part of the LFs, and hence the response functions, are concerned.
The real parts instead  are altered by an extra  contact term
that matters in the response when higher orders (say, a RPA series) 
are accounted for.

In attempting to account for the different off-shell behaviour
one meets a proliferation of complicated and mostly irrelevant terms. 
Thus here we confine ourselves to consider only the vertices 
${\cal O}=I$  ($\sigma$-meson absorption),  $\gamma_5$ and $\Not q \gamma_5$ 
(pion absorption within the pseudoscalar and the pseudovector coupling).
This last, at variance of the pseudoscalar coupling, correctly
describes the $\pi^0$ suppression in the photo-production process and
respects the chiral limit, owing to a further contact term
added to the pseudoscalar vertex.
Notice that the vertices containing a $\gamma_5$ are derived
from the corresponding parity-conserving ones (up to, eventually, a sign) 
by replacing $m$ with $-m$.

For the 0-index LFs Eq.~\eqref{eq:A862} 
applies and we only need to specify $A^S$ and $\lambda^S_{mn}$
in the various cases.

Considering first a scalar probe, the function $f_{ss}$  reads
\begin{equation}
  \label{eq:A123}
  \begin{split}
    f_{ss}(p,q)&=\Tr (\Not p+\Not q+M)(\Not p+m)=4(p^2+p\cdot q+Mm)\\
    &=2\D^*(p+q)-2\D(p)+2Q_+~,
  \end{split}
\end{equation}
where use has been made of  the identity \eqref{eq:A710} in the second 
line.
Thus we get $A^S(q^2)=2Q_+$ and 
$\lambda^S_{10}=2$. The term $2\D(p)$ is cancelled by the
renormalisation when studying the resonance-hole case, but it survives
in the nucleon-hole one. In conclusion
\begin{equation}
  \label{eq:A435}
  \Pi^{*(0)}_{ss}(M;\q,q_0)=2 T_{S+}^{*[1,0]}+2Q_+\Upsilon^{*[0]}_+(M;\q,q_0)~.
\end{equation}

The other  vertices are decoupled
from the identity by parity conservation.
For the pseudoscalar coupling ($\gamma_5$) we find
\begin{equation}
  \label{eq:D435}
  \Pi^{*(0)}_{psps}(M;\q,q_0)=-\Pi^{*(0)}_{ss}(M;\q,q_0)\biggm|_{m\to-m}~.
\end{equation}
For the pseudovector coupling, introducing the notations
\begin{equation}
  \label{eq:A805}
  M_T=M+m\qquad\qquad\qquad\delta m=M-m~,
\end{equation}
we find 
$A^S=2M_T^2Q_-$ and $\lambda^S_{10}=2(\delta m M_T-q^2\rho)$, 
$\lambda^S_{20}=2$.
As already outlined, in going from the pseudoscalar to the pseudovector 
LF the coefficient $A^S$ is multiplied by the expected factor 
$M_T^2$, but the contact term does not. 

The pseudovector coupling conserves the axial current 
(or alternatively the existence of the Goldstone boson).
Covariance would entail $\Pi^{*}_{PV}\sim q^2$,
but, since it is broken, the Goldstone theorem only requires
\begin{equation}
  \label{eq:A699}
  \lim_{\substack{q_0\to0\\\q\to0}}\Pi^{*}_{PV}(M;\q,q_0)=0~.
\end{equation}
Now we observe that 
\begin{equation}
  \label{eq:A701}
  \Upsilon_+^{[0]}(M;0,0)=2U_+^{[0]}(M;0,0)=2\intt{p}\frac{1}{2E_p}
  \frac{\theta(k_F-\p)}{m^2-M^2}=-\frac{2}{M^2-m^2}\T^{[0]}
\end{equation}
and  it is easily verified that
 the contact terms are tailored in such a way to exactly cancel 
$A^S\Upsilon_+^{[0]}$ in the limit
$q\to0$.

Finally the mixed pseudoscalar-pseudovector LF also
exists: it has $A^S=2M_TQ_-$ and $\lambda^S_{10}=2\delta m$ while
the pseudovector-pseudoscalar function has the opposite sign.

The nucleon-hole case  must be considered aside because of the different 
structure of the contact terms. For example, the scalar-scalar response,
owing to \eqref{eq:A123} leads to the manifestly vanishing contact term
$$-2i\intq{p}\S_m(p)+2i\intq{p}\S_m(p+q)~.$$
It is found that the only existing
contact term pertains to the pseudovector-pseudovector case and is given by
$4q^2\T^{[0]}$.

\subsection{The scalar-vector interference}
\label{sec:7c}

In infinite nuclear matter a scalar probe can be transformed, 
through a resonance (nucleon)-hole propagator, into a vector one. 
Thus we shall consider, as before, the 
scalar-type vertices $I$ and $\gamma_5$.
Concerning the vector-like vertices, 24 independent
currents exist, 12 of them parity conserving  and 12 parity-violating
(see, {\it e.g.}, Ref.~\cite{Be-68-B}).
However, currents embodying a $q^\mu$ (that can be extracted out of the
integral) times a (pseudo-)scalar structure reduce to
a 0-index LF, already handled in Sec.~\ref{sec:7.1},
times  $q^\mu$. Further, neglecting contact terms,
$\Not p$  and $\Not q$ are redundant and the Gordon identity
\begin{multline}
  \label{eq:A905}
  p^\mu=-\frac{1}{2}q^\mu+\frac{1}{4}(\not p+\not q-M)\gamma^\mu
  +\frac{1}{4}\gamma^\mu(\not p-m)\\+\frac{1}{4}(M+m)\gamma^\mu
  -\frac{i}{4}\sigma^{\mu\nu}q_\nu
\end{multline}
allows us to express the current $p^\mu$ in terms of the usual currents 
$\gamma^\mu$ and $\sigma^{\mu\nu}q_\nu$ {\em only on the mass shell}, 
while the  off-mass-shell extension of the currents remains unpredictable.

Disregarding the huge variety of contact terms and $q^\mu$, 
only four independent currents survive, and they may be forced to be
conserved by adding  some suitable terms proportional to $q^\mu$.
They read
\begin{subeqnarray}
  \label{eq:A875}
  j_1^\mu&=&\gamma^\mu-\frac{\Not q q^\mu}{q^2}\\
  j_2^\mu&=&\frac{i}{2m}q_\nu \sigma^{\mu\nu}\\
  j_{3,4}^\mu&=&j_{1,2}^\mu\gamma_5~.
\end{subeqnarray}
Furthermore $j_4^\mu$  is expressible in terms of the other three
currents by exploiting the charge conjugation symmetry.

Since  all these current are conserved, only the first term of 
Eq.~\eqref{eq:A327} is required.
The functions $A^V_{xy}(q^2)$ are listed, with a self-explanatory notation,
in  Table \ref{tab:C1}.
\begin{table}[ht]
  \begin{center}
    \leavevmode
    \begin{tabular}{||c||c|c|c|c||}
      \hhline{|t=====|}
      ~&$I\to j_1^\mu$&$I\to j_2^\mu$&$\gamma_5\to j_3^\mu$
      &$\gamma_5\to j_4^\mu$\cr
      \hhline{||=#====||}
      \rb{$A^{(1)}_{xy}(q^2)$}&\rb{$4M_T$}&
      \rb{$\dfrac{2}{m}q^2$}&\rb{$4\delta m$}&\rb{$-\dfrac{2}{m}q^2$}\cr
      \hhline{|b=====|}
    \end{tabular}
    \caption{The function $A^V_{xy}(q^2)$ for the 
      scalar-vector Lindhard function}
    \label{tab:C1}
  \end{center}
\end{table}

\subsection{The 2-indices response}
\label{sec:7b}

\begin{table}[ht]
  \begin{minipage}{60mm}
    \begin{center}
      \leavevmode
      \begin{tabular}{||c||c|c||}
        \hhline{|t===|}
        ~&$j_1^\mu$&$j_2^\mu$\cr
        \hhline{||=#==||}
        \rb{$j_1^\nu$}&\rb{$-\dfrac{2Q_-}{q^2}$}&\rb{$\dfrac{M_T}{mq^2}Q_-$}
       \cr
        \hhline{||-||--||}
        \rb{$j_2^\nu$}&\rb{$-\dfrac{M_T}{mq^2}Q_-$}
        &\rb{$\dfrac{M_T^2}{2m^2q^2}Q_-$}\cr
        \hhline{|b===|}
      \end{tabular}
      \caption{The functions $A^S(q^2)$ for the vector-vector 
        parity conserving 
        currents}
      \label{tab:A7}
    \end{center}
  \end{minipage}
  \hskip1cm
  \begin{minipage}{60mm}
    \begin{center}
      \leavevmode      
      \begin{tabular}{||c||c|c||}
        \hhline{|t===|}
        ~&$j_1^\mu$&$j_2^\mu$\cr
        \hhline{||=#==||}
        \rb{$j_1^\nu$}&\rb{$8$}&
        \rb{0}\cr
        \hhline{||-||--||}
        \rb{$ j_2^\nu$}&\rb{0}
        &\rb{$2\dfrac{q^2}{m^2}$}\cr
        \hhline{|b===|}
      \end{tabular}
      \caption{The functions $A^{LT}(q^2)$ for the vector-vector 
        parity conserving currents}
      \label{tab:A9}
    \end{center}
  \end{minipage}
\end{table}

We consider now the vector-vector response and distinguish between three
different sets of LFs. 
\begin{enumerate}
\item First we examine the parity conserving-parity conserving 
LFs (set up by the conserved currents $j_1^\mu$ and $j_2^\mu$), which
are symmetric tensors. Thus only the
functions $\Pi^{*L}$ and $\Pi^{*T}$ are required, that in turn need the
knowledge of $A^S$, $A^{LT}$  and of the corresponding $\lambda$'s,
according to
\eqref{eq:A878} and \eqref{eq:A567}.
The $A^S$ are summarised in 
Table  \ref{tab:A7} and the $A^{LT}$ in \ref{tab:A9}. The contact terms
$\lambda^S_{10}$
pertaining to $\Pi^{*S}$ are displayed in Table \ref{tab:B3}. The other
contributions, namely the $\lambda^{LT}_{i0}(q^2)$ are all vanishing.
Instead a coefficient $\lambda^S_{20}$ survives for the $j_2^\mu j_2^\nu$
case and the relation $\lambda^S_{20}(j_2^\mu j_2^\nu)=1/(2m^2q^2)$ holds.

\begin{table}[ht]
  \begin{minipage}{63mm}
    \begin{center}
      \leavevmode
      \begin{tabular}{||c||c|c||}
        \hhline{|t===|}
        ~&$j_1^\mu$&$j_2^\mu$\cr
        \hhline{||=#==|}
        \rb{$j_1^\nu$}&\rb{$-\dfrac{2}{q^2}$}
        &\rb{$\dfrac{\delta m}{mq^2}$}\cr
        \hhline{||-||--||}
        \rb{$j_2^\nu$}&\rb{$-\dfrac{\delta m}{mq^2}$}
        &\rb{$-\dfrac{q^2\rho-M_T\delta m}{2m^2q^2}$}\cr
        \hhline{|b===|}
      \end{tabular}
      \caption{The contact terms
        $\lambda^S_{10}(q^2)$  for the vector-vector 
        parity conserving 
        currents}
      \label{tab:B3}
    \end{center}
  \end{minipage}
  \hskip1cm
  \begin{minipage}{63mm}
    \begin{center}
      \leavevmode
      \begin{tabular}{||c||c|c||}
        \hhline{|t===|}
        ~&$j_1^\mu$&$j_2^\mu$\cr
        \hhline{||=#==|}
        \rb{$j_3^\nu$}&\rb{$4i$}
        &\rb{$-2i\dfrac{M_T}{m}$}\cr
        \hhline{||-||--||}
        \rb{$j_{4}^\nu$}&\rb{$-2i\dfrac{\delta m}{m}$}&
        \rb{$i\dfrac{M_T\delta m}{m^2}$}\cr
        \hhline{|b===|}
      \end{tabular}
      \caption{The functions 
        $A^{*V}(q^2)$  for the vector-vector 
        parity conserving-parity  violating 
        currents}
      \label{tab:B4}
    \end{center}
  \end{minipage}
\end{table}

\item Next we consider parity conserving-parity violating 
LFs (currents $j_1^\mu$ and $j_2^\mu$ at the incoming vertex,
$j_3^\mu$ and $j_4^\mu$ at the outgoing one). Here the tensors are 
antisymmetric and Eq.~\eqref{eq:A685} applies with  the second term only,
namely
$$q_\lambda \N^\sigma g_{\sigma\rho} 
  \epsilon^{\mu\nu\rho\lambda}\Pi^{*V}_{xy-}(M;\q,q_0)~,$$
with $\Pi^{*V}_{xy-}$ defined by  Eq.~\eqref{eq:A821}. The required
functions $A^{*V}(q^2)$ are displayed in Table \ref{tab:B4}. 
No contact term exists. The case $j_2^\mu j_4^\nu$
would provide a non-vanishing function $\lambda^V_{10}$, but actually
$T^{[1,0]}_{x-}=0$.
Finally if we reverse the vertices the following relation holds:
\begin{equation}
  \label{eq:A733}
  \Pi^{(2)\mu\nu}_{j_n^\mu j_m^\nu}=(-1)^{m+1}
  \Pi^{(2)\mu\nu}_{j_m^\mu j_n^\nu}~,~~~~~m=1,2~,~~~~~n=3,4~.
\end{equation}

\item 
The parity violating-parity violating LFs are derived from
the parity conserving-parity conserving ones through the simple
relations
\begin{subeqnarray}
  \label{eq:E001}
  \Pi^*_{j_3,j_{3,4}}&=&\Pi^*_{j_1,j_{1,2}}|_{m\to-m}~,\\
  \Pi^*_{j_4,j_{3,4}}&=&-\Pi^*_{j_2,j_{1,2}}|_{m\to-m}~.
\end{subeqnarray}
\end{enumerate}

Again the nucleon-hole case differs from the above only for the contact terms, 
because a direct evaluation shows that the various $\Q$ are just
the limits of the $\Q^*$ for $M\to m$. Only one contact term
exists for the case $j_1^\mu j_1^\nu$, with $\lambda_{10}^{S}=-2/q^2$.

\section{The spin 3/2 resonances}
\label{sec:9}

We consider now the excitation of a nucleon to a spin 3/2 resonance 
(specifically the $\Delta(1232)$ ), assumed to be stable. The 
resonance is described by a vector-spinor field $\psi_\mu$ obeying the
Rarita-Schwinger equations
\begin{subeqnarray}
  \label{eq:A915}
  (i\Not\partial-m)\psi_\mu&=&0\\
  \gamma^\mu\psi_\mu&=&0\\
  \partial^\mu\psi_\mu&=&0
\end{subeqnarray}
(the last line is, more properly, a constraint)
which can be deduced from the Lagrangian~\cite{MoCa-56}
\begin{multline}
  \label{eq:A918}
  {\cal L}=\bar\psi_\mu\Bigl\{
  (i\Not\partial-M)g^{\mu\nu}+
  i\omega(\gamma^\mu\partial^\nu+\partial^\mu\gamma^\nu)\\
  +\frac{i}{2}\left(3\omega^2+2\omega+1\right)\gamma^\mu\Not\partial\gamma^\nu
  +\left(3\omega^2+3\omega+1\right)M\gamma^\mu\gamma^\nu\Bigr\}\psi_\nu~,
\end{multline}
$\omega$ ($\not=-1/2$) being a free parameter.
Each value of $\omega$ leads to the equations of motion \eqref{eq:A915},
but does not prevent the occurrence of a 
spin 1/2 component in the $\omega$-dependent vector-spinor $\psi_\mu$.
Thus, to rule out these unwanted components, one usually introduces
the projection operator on the spin 3/2 space, which reads
(in momentum space)
\begin{equation}
  \label{eq:A920}
  P_{3/2}^{\mu\nu}=g^{\mu\nu}-\frac{1}{3}\gamma^\mu\gamma^\nu-\frac{1}{3p^2}
  \left(\Not p\gamma^\mu p^\nu+p^\mu\gamma^\nu\Not p\right)~,
\end{equation}
or, sometimes, its on-shell reduction
\begin{equation}
  \label{eq:A940}
  P_{3/2}^{\mu\nu}\rightarrow P_{3/2}^{\mu\nu}\bigm|_{p^2\to M^2}~.
\end{equation}
The most common choices are $\omega=-1/3$, that leads to the Rarita-Schwinger
result, and $\omega=-1$, that corresponds to the Lagrangian
\begin{equation}
  \label{eq:A919}
  {\cal L}=\bar\psi_\mu\left\{-\epsilon^{\mu\nu\lambda\rho}
  \gamma^5\gamma_\lambda\partial_\rho-iM\sigma^{\mu\nu}\right\}\psi_\nu~.
\end{equation}
Another Lagrangian has been recently proposed, namely~\cite{KiNa-04}
\begin{equation}
  \label{eq:A921}
  {\cal L}=\bar\psi_\mu\left\{-\epsilon^{\mu\nu\lambda\rho}
  \gamma^5\gamma_\lambda\partial_\rho-Mg^{\mu\nu}\right\}\psi_\nu
\end{equation}
in order to solve the so-called Velo-Zwanziger disease~\cite{VeZw-69,VeZw-69a}.
However we do not discuss such a Lagrangian here, as it describes a resonance
propagating in an external electro-magnetic field
with the (minimally coupled) $\Delta\Delta\gamma$ vertex, while we 
only consider non-minimal $N$-$\Delta$ transitions. 
Furthermore, the proposal \eqref{eq:A921} 
is seriously plagued by the occurrence
of a pole at $p^2=M^2/4$, as the evaluation of the propagator (the inverse of
$\Gamma^{\mu\nu}$) shows.

Sticking to the more sound form \eqref{eq:A918}, observe that different
values  of  $\omega$ not only alter the mixing between 3/2 and 1/2 spin,
but also affect  the off-shell behaviour of the $\Delta$-hole propagator, that
in fact reads
\begin{multline}
  \label{eq:A925}
  \Delta^{\mu\nu}=\frac{\Not p+M}{p^2-M^2}
  \left\{g^{\mu\nu}-\frac{1}{3}\gamma^\mu\gamma^\nu-
    \frac{1}{3M}(\gamma^\mu p^\nu-\gamma^\nu p^\mu)-\frac{2}{3M^2}
    p^\mu p^\nu\right\}\\
  +\frac{1+\omega}{6M^2(1+2\omega)^2}\Bigl\{
  (1+3\omega)(\gamma^\mu p^\nu+\gamma^\nu p^\mu)
  +(1+\omega)(\gamma^\mu p^\nu-\gamma^\nu p^\mu)\\
  -2M\omega\gamma^\mu\gamma^\nu
  +(1+\omega)\Not p\gamma^\mu\gamma^\nu\Bigr\}
\end{multline}
Remarkably the choice $\omega=-1$ cancels in \eqref{eq:A925}
all the terms with no analytic structure, which would otherwise contribute
to the contact terms in the $\Delta$-hole LF.

\subsection{The 0-index $\Delta$-hole Lindhard functions}
\label{sec:9.1}

Here the vertices have the form $\bar \psi j^\mu\psi_\mu$ or
$\bar \psi j^\mu P_{3/2}^{\mu\nu}\psi_\nu$, where $j^\mu$ could 
be taken from Eq.~\eqref{eq:A875} plus the non-conserved $q^\mu$ 
and $\gamma_5q^\mu$. Clearly
\begin{equation}
  \label{eq:A937}
  f_{xy}(p,q)=\Tr(\Not p+m){j_x}_\mu\Delta^{\mu\nu}{j_y}_\nu
\end{equation}
or, alternatively,
\begin{equation}
  \label{eq:A938}
  f_{xy}(p,q)=\Tr(\Not p+m){j_x}_\mu P_{3/2}^{\mu\lambda}
  \Delta^{\lambda}_{\ \rho}P_{3/2}^{\rho\nu}{j_y}_\nu~.
\end{equation}
However, the vector $\gamma^\mu$ is constrained by (\ref{eq:A915}b)
so that this current, when contracted with $\Delta^{\mu\nu}$,
is vanishing on shell, does not develop an imaginary part and consequently it
can only contribute to the contact terms. The same occurs
for $\gamma^\mu\gamma_5$. 
Furthermore $\sigma^{\mu\lambda}q_\lambda$ can be replaced
by $q^\mu$ because
$$\sigma^{\mu\nu}q_\nu\psi_\mu=iq^\mu\psi_\mu~.$$ 
Thus only the two currents
\begin{subeqnarray}
  \label{eq:A939}
  j_1(N\Delta)^\mu&=&q^\mu\\
  j_2(N\Delta)^\mu&=&q^\mu\gamma_5
\end{subeqnarray}
actually matter, at least for the part carrying analytic structure.
Since we are dealing with a 0-index LF, the structure
is given by \eqref{eq:A862}.
With self-explanatory notations the non-vanishing $A^{S}$ are found to be
\begin{equation}
  \label{eq:A948}
  A^{S}_{j_1^\mu j_1^\nu}=-\frac{Q_-Q_+^2}{3M^2}~,~~~~~
  A^{S}_{j_2^\mu j_2^\nu}=\frac{Q_-^2Q_+}{3M^2}~.
\end{equation}
Concerning the contact terms, we will not give a detailed list for all the 24
 currents, because they all explicitly
depend upon $\omega$ and display a double pole at $\omega=-1/2$:
hence they can diverge and the real part of the LFs becomes unpredictable. 
As a consequence, the RPA series based on the $\Delta$-hole excitation
(and, similarly, any calculation
beyond the bare Free Fermi Gas) becomes unreliable.

The same happens if we use the expression \eqref{eq:A938}
taking however the projection operator in the form \eqref{eq:A940}.
We get indeed Eqs.~\eqref{eq:A948} for the functions $A^{S}$, but
again the contact terms display a double pole in $\omega$.

Finally we can take the projection operator in the form \eqref{eq:A920}.
Now the expressions \eqref{eq:A948} are again valid, but the contact terms
are independent of $\omega$ and, furthermore, they
do not change in replacing $q^\mu$ with $i\sigma^{\mu\nu}q_\nu$.
They display however a factor $(p+q)^{-2}$
coming from the projection operator Eq.~\eqref{eq:A920}.
For instance, in the current $j_1{(N\Delta)}^\mu$ they take the form
\begin{multline*}
  \intt{p}\frac{1}{2E_p}\Biggl\{
  \frac{(m^2-q^2)^2(m^2+2m M-q^2)}{3M^2(p+q)^2}\Bigm|_{p^2=m^2}
  \\+\frac{1}{3}(2q^2-2p\cdot q-M^2-2m M)\Biggr\}\theta(k_F-\p)
  +(q_0\longleftrightarrow -q_0)~.
\end{multline*}
Here the first term is just the function 
$U^{*[0]}_{\rm rel}$ evaluated at $M=0$, thus explaining the introduction
of the factor  $B_{xy}^S(q^2)\Upsilon^{*[0]}_\pm(M=0;\q,q_0)$ in 
Eq.~\eqref{eq:A862}.
Explicitly the case $(j_1^\mu j_1^\nu)$ requires
\begin{equation}
  \label{eq:V021}
  B^S_{j_1^\mu j_1^\nu}=\frac{(m^2-q^2)^2(m^2+ 2m M-q^2)}{3M^2}
\end{equation}
and furthermore 
$\lambda^S_{10}=(m^2-2m M-2M^2+3q^2)/3$ and $\lambda^S_{20}=-1/3$.
Finally,
\begin{equation}
  \label{eq:V022}
  \Pi^{*(0)}_{j_2^\mu j_2^\nu}=-\Pi^{*(0)}_{j_1^\mu j_1^\nu}\bigm|_{m\to-m}~.
\end{equation}

\subsection{The 1-index function}
\label{sec:9.2}

Now we consider the case of the transition from a scalar to a vector term,
which leads to a 1-index LF. Here, besides the vectors 
discussed in the previous subsection, we also need a tensor
operator yielding a vector when contracted with the $\Delta$ propagator.
Again one can set up a vast amount  of tensors: we limit ourselves to 
those which give LFs with the same analytic structure, ignoring 
extra contact terms. 

We are thus left with eight possible currents, namely
\begin{equation}
  \label{eq:A927}
  j^\nu_X=\overline\psi \Gamma^{\mu\nu}_X\psi_\mu+{\rm h.~c.}
\end{equation}
with
\begin{subeqnarray}
  \label{eq:A964}
  \Gamma^{\mu\nu}_{M(V)}&=&-\frac{3}{2}\frac{\mu+1}{Q_+}
  \epsilon^{\mu\nu\alpha\beta}p_\alpha q_\beta\\
  \Gamma^{\mu\nu}_{E(V)}&=&-\Gamma^{\mu\nu}_{M(V)}-i\frac{3}{2}
  \frac{\mu+1}{Q_+Q_-}4\epsilon^{\mu\sigma\alpha\beta}p_\alpha q_\beta
  \epsilon^\nu_{\;\sigma\gamma\delta}p^\gamma q^\delta\gamma_5\\
  \Gamma^{\mu\nu}_{C(V)}&=&-i\frac{3}{2}
  \frac{\mu+1}{Q_+Q_-}2q^\mu\left(q^2 p^\nu-p\cdot q q^\nu\right)\gamma_5\\
  \Gamma^{\mu\nu}_{M(A)}&=&-i\frac{3}{2}\frac{\mu-1}{Q_-}\left[-2\gamma_5
    \epsilon^{\mu\nu\alpha\beta}p_\alpha q_\beta-\frac{2}{Q_+}
    \epsilon^{\mu\sigma\alpha\beta}p_\alpha q_\beta
    \epsilon^\nu_{\;\sigma\gamma\delta}p^\gamma q^\delta\right]\\
  \Gamma^{\mu\nu}_{E(A)}&=&-i\frac{3}{2}\frac{\mu-1}{Q_+Q_-}
  2\epsilon^{\mu\sigma\alpha\beta}p_\alpha q_\beta
  \epsilon^\nu_{\;\sigma\gamma\delta}p^\gamma q^\delta\\
  \Gamma^{\mu\nu}_{C(A)}&=&i\frac{3}{2}
  \frac{\mu-1}{Q_+Q_-}2q^\mu\left(q^2 p^\nu-p\cdot q q^\nu\right)\\
  \Gamma^{\mu\nu}_{S(V)}&=&g^{\mu\nu}\gamma_5\\
  \Gamma^{\mu\nu}_{S(A)}&=&g^{\mu\nu}
\end{subeqnarray}
(here $\mu=M/m$).
We have followed in the above the work of Devenish et al.~\cite{DeEiKo-76}:
these author show that, for a transition from a nucleon to a higher spin
resonance (not necessarily 3/2), 
only three conserved currents enter the parity-conserving sector
and as many the parity-violating one. In the low momentum regime
tensors associated to these six currents correspond to the multipoles 
M1, E2, C2, M2, E1, C1
(the first three parity conserving, the other parity violating).
We have added two other currents, which are not conserved, thus exhausting
all the possibilities.

We see that again  LFs
with only one vector index exists, but they only occur 
between a Coulomb multipole and  the non-conserved currents \eqref{eq:A939}.
The case $(j_2^\mu \Gamma_{C(V)})$ 
contains only a $t^\mu$ in the integrand, thus the expression \eqref{eq:A821} 
applies (first term in \eqref{eq:A327}) with
\begin{subeqnarray}
  \label{eq:A618}
  A^V_{j_2^\mu \Gamma_{C(V)}}&=&
  -i\frac{M_T}{mM^2}Q_-\\
  B^V_{j_2^\mu \Gamma_{C(V)}}&=&i\frac{M_T}{mM^2}\frac{
    (m^2-q^2)^2(m^2-2mM-q^2)q^2}{Q_+Q_-}\\
  \lambda^V_{10}&=&\frac{iM_Tq^2}{m}\frac{m^2+2mM-2M^2+3q^2}{Q_+Q_-}\\
  \lambda^V_{20}&=&-\frac{iM_Tq^2}{mQ_+Q_-}~.
\end{subeqnarray}
The case $(j_2^\mu \Gamma_{S(V)})$ requires the full expression \eqref{eq:A327}
and we find (since $T^{[1,0]}_{S-}=0$)
\begin{subeqnarray}
  \label{eq:A619}
  A^V_{j_2^\mu \Gamma_{S(V)}}&=&
  \frac{2Q_-}{3M^2}(q^2+M^2-m^2)\\
  B^V_{j_2^\mu \Gamma_{S(V)}}&=&\frac{2}{3M^2}(m^2-2mM-q^2)(m^2-q^2)\\
  \lambda^V_{10}&=&\frac{2}{3}\\
  A^S_{j_2^\mu \Gamma_{S(V)}}&=&\frac{Q_-^2Q_+q^2}{3M^2}\\
  B^S_{j_2^\mu \Gamma_{S(V)}}&=&-
  \frac{1}{3M^2q^2}(m^2-2mM-q^2)(m^2-q^2)^2\\
  \lambda^S_{20}&=&\frac{1}{3q^2}~.
\end{subeqnarray}
The LFs with opposite parity simply obtain as
\begin{equation}
  \label{eq:A981}
  \Pi^{*(1)\mu}_{j_1,\Gamma_{x(A)}}(\q,q_0)=-
  \Pi^{*(1)\mu}_{j_2,\Gamma_{x(V)}}(\q,q_0)\bigm|_{m\to-m}~.
\end{equation}

\subsection{The 2-indices functions}
\label{sec:9.3}

\subsubsection{Parity conserving-parity conserving Lindhard functions}

We consider first of all the couplings $M1$, $E2$ and $C2$ (parity conserving)
in both vertices. These currents being conserved, we can directly apply
Eqs.~\eqref{eq:A878} and \eqref{eq:A567}. Hence the functions
$A_{xy}^S$, $A_{xy}^{LT}$ and the corresponding $\lambda$'s are needed.
Remarkably, $A_{xy}^S$ and $A_{xy}^{LT}$ are diagonal with respect to the 
channel indices and are quoted in Tables \ref{tab:X1} and
\ref{tab:X2} respectively.
\begin{table}[ht]
  \begin{minipage}{62mm}
    \begin{center}
      \leavevmode
      \begin{tabular}{||c|c|c||}
        \hhline{|t===|}
        $\Gamma_{M(V)}$&$\Gamma_{E(V)}$&$\Gamma_{C(V)}$\cr
        \hhline{||===||}
        \rb{$\dfrac{3M_T^2}{4m^2 q^2}Q_-$}&
        \rb{$\dfrac{9M_T^2}{4m^2 q^2}Q_-$}
        &\rb{$0$}\cr
        \hhline{|b===|}
      \end{tabular}
      \caption{The $A^S(q^2)$ functions for the vector-vector $\Delta$-$N$
        parity conserving 
        currents}
      \label{tab:X1}
    \end{center}
  \end{minipage}
  \hskip1cm
  \begin{minipage}{62mm}
    \begin{center}
      \leavevmode      
      \begin{tabular}{||c|c|c||}
        \hhline{|t===|}
        $\Gamma_{M(V)}$&$\Gamma_{E(V)}$&$\Gamma_{C(V)}$\cr
        \hhline{||===||}
        \rb{$\dfrac{3M^2_Tq^2}{m^2Q_+}$}&
        \rb{$\dfrac{9M_T^2q^2}{m^2Q_+}$}
        &\rb{$-\dfrac{3M_T^2q^4}{m^2M^2Q_+}$}\cr
        \hhline{|b===|}
      \end{tabular}
      \caption{The $A^{LT}(q^2)$ functions for the vector-vector $\Delta$-$N$
        parity conserving
        currents}
      \label{tab:X2}
    \end{center}
  \end{minipage}
\end{table}
Instead, a contact term coupling the multipoles $M1$ and $E2$ exists
(it is reported in Appendix \ref{sec:appE} together with all the other contact
terms).

Both the multipoles $M1$ and $E2$ display the structure
of Eqs.~\eqref{eq:A878} and \eqref{eq:A567}  with, however,
$B^x=0$, while in the Coulomb multipole $C2$ the contribution
proportional to $\Pi^{*S}$ is absent.
$ A^{LT}$ is given in Table \ref{tab:X2} 
and furthermore
\begin{equation}
  B^{LT}_{C(V)C(V)}(q^2)=\frac{3M_T^2q^4}{m^2M^2Q_+^2Q_-^2}
  (m^2-q^2)^2(m^2-2mM-q^2)~.
\end{equation}
Observe also that in this sector $\Pi^{*(2)\mu\nu}_{xy}=\Pi^{*(2)\mu\nu}_{yx}$.

\subsubsection{Parity conserving-parity violating Lindhard functions}

This kind of LF is antisymmetric in the indices $\mu$, $\nu$
and takes the form
$$q_\lambda \N^\sigma g_{\sigma\rho} 
  \epsilon^{\mu\nu\rho\lambda}\Pi^{*V}_{xy-}(M;\q,q_0)$$
(see Eq.~\eqref{eq:A821}). We thus need to specify the functions $A^V_{xy}$
and the contact terms. The only non-vanishing $A^V_{xy}$ are
\begin{equation}
  A^V_{E(V)M(A)}(q^2)=3
  A^V_{M(V)E(A)}(q^2)=-\frac{9iM_T\delta m}{2m^2}
\end{equation}
and for the relevant contact terms (not $\lambda^V_{10}$ 
$T^{[1,0]}_{V-}=0$) we refer the reader to Appendix \ref{sec:appE}.
Moreover,
\begin{equation}
  \label{eq:A119}
  \Pi^{0\mu\nu}_{x(A)y(V)}=-\Pi^{0\mu\nu}_{x(V)y(A)}
\end{equation}
(with $x,y=M,E,C$). 

\subsubsection{Parity violating-parity violating Lindhard functions}

The functions $A^S$, $A^{LT}$ and  $B^{LT}$ 
follow from the parity conserving-parity conserving
case with the replacements
\begin{multline}
  A^x_{E(A)E(A)}(q^2)=-A^x_{M(V)M(V)}(q^2)\bigm|_{m\to -m}~,\\
  A^x_{M(A)M(A)}(q^2)=-A^x_{E(V)E(V)}(q^2)\bigm|_{m\to -m}~,\\
  A^{LT}_{C(A)C(A)}=-A^{LT}_{C(V)C(V)}\bigm|_{m\to -m}~,
  B^{LT}_{C(A)C(A)}=-B^{LT}_{C(V)C(V)}\bigm|_{m\to -m}~.
\end{multline}
As for the parity conserving-parity conserving
case the rule (involving contact terms) 
$\Pi^{*(2)\mu\nu}_{xy}=\Pi^{*(2)\mu\nu}_{yx}$ holds.
The contact terms (see Appendix~\ref{sec:appE})
have a quite involved structure.

\subsubsection{Lindhard functions involving non-conserved currents}

A LF having the first vertex $M(V)$ or $E(V)$ and the second
 $S(V)$ (this last corresponds to a non-conserved current)
is non-vanishing and has the structure of Eq.~\eqref{eq:A881} (a symmetric
LF obeying the current conservation law).
Thus it can be expressed in terms of Eqs.~\eqref{eq:A880} with
\begin{equation}
  \label{eq:V001}
  3A^S_{M(V)S(V)}=A^S_{E(V)S(V)}=\frac{3iM_TQ_-}{2mq^2}
\end{equation}
and
\begin{equation}
  \label{eq:V002}
  3A^{LT}_{M(V)S(V)}=A^{LT}_{E(V)S(V)}=\frac{6iM_Tq^2}{mQ_+}~.
\end{equation}
The same happens for $\Pi^{*(2)\mu\nu}_{E(A)S(A)}$
with
\begin{equation}
  \label{eq:V016}
  A^S_{E(A)S(A)}=-i\frac{\delta m Q_+}{mq^2}~,~~~~~~~A^{LT}_{E(A)S(A)}
  =-4i\frac{\delta mq^2}{mQ_-}~,
\end{equation}
while $\Pi^{*(2)\mu\nu}_{M(A)S(A)}$ only displays contact terms
(see Appendix \ref{sec:appE}).

The  LF $\Pi^{*(2)\mu\nu}_{C(V)S(V)}$ has a different 
structure: it contains a symmetric current-conserving term, as in 
Eq.~\eqref{eq:A881}, with $A^S_{C(V)S(V)}=0$ and 
\begin{subeqnarray}
  \label{eq:V004}
  A^{LT}_{C(V)S(V)}&=&\frac{2iM_Tq^2}{mM^2Q_+}\left[m^2-M^2-q^2\right]\\
  B^{LT}_{C(V)S(V)}&=&-\frac{2iM_Tq^2}{mM^2Q_+Q_-}(m^2-q^2)(m^2-2mM-q^2)
\end{subeqnarray}
plus the following term proportional to $q^\nu$
\begin{multline}
  \label{eq:V005}
  \N^\mu q^\nu\Bigl[-\frac{iM_TQ_-}{mM^2}\Q^{*V}_-\\+\frac{iM_T}{mM^2Q_+Q_-}
    (m^2-q^2)^2(m^2-2mM-q^2)\Q^{*V}_-(M=0)\\
    +\sum_m \lambda^V_{m0}
      T^{*[m-1,0]}_{V-}(\q,q_0)\Bigr]~,
\end{multline}
with the contact terms given in Appendix \ref{sec:appE}. The symmetry relation
\begin{equation}
  \label{eq:V018}
  \Pi^{*(2)\mu\nu}_{C(A)S(A)}=-\Pi^{*(2)\mu\nu}_{C(V)S(V)}\bigm|_{m\to-m}
\end{equation}
holds.

The functions with initial vertex $M(A)$ and $E(A)$ and final vertex $S(V)$,
as well as $\Pi^{*(2)\mu\nu}_{M(V)S(A)}$ and $\Pi^{*(2)\mu\nu}_{S(V)S(A)}$ 
are antisymmetric and are given by the second term of Eq.~\eqref{eq:A685},
namely 
$$q_\lambda \N^\sigma g_{\sigma\rho} 
  \epsilon^{\mu\nu\rho\lambda}\Pi^{*V}_{xy-}(M;\q,q_0)~,$$
with
\begin{align}
  \label{eq:V007}
  A^V_{M(A)S(V)}&=3 A^V_{E(A)S(V)}=-3\frac{\delta m}{m}~,\\
  \label{eq:V015}
  A^V_{M(V)S(A)}&=\frac{2M_T}{m}\\
\intertext{and}
  \label{eq:V012}
  A^V_{S(V)S(A)}&=\frac{4i}{3}~,
\end{align}
while $\Pi^{*(2)\mu\nu}_{E(V)S(A)}$
has only contact terms and $\Pi^{*(2)\mu\nu}_{C(A)S(V)}=
\Pi^{*(2)\mu\nu}_{C(V)S(A)}=0$.

Next we consider the symmetric function $\Pi^{*(2)\mu\nu}_{S(V)S(V)}$. 
Since the current is not conserved, it is contributed to by
all the terms in Eq.~\eqref{eq:A663}. Thus it displays a conserved part, 
which has
\begin{subeqnarray}
  \label{eq:V009}
  A^{S}_{S(V)S(V)}&=&-\frac{4Q_-}{3q^2}\\
  A^{LT}_{S(V)S(V)}&=&\frac{4Q_-}{3M^2}\\
  B^{LT}_{S(V)S(V)}&=&\frac{4(q^2-m^2+2mM)}{3M^2}
\end{subeqnarray}
plus a non-conserved one (last two terms in \eqref{eq:A663}).
The latter requires the knowledge of $\Pi^{*V}_{S(V)S(V)-}$ -
in turn fixed by (see Eq.~\eqref{eq:A821})
\begin{eqnarray}
  \label{eq:V010}
  A^{V}_{S(V)S(V)}&=&\frac{2Q_-}{3M^2q^2}(M^2-m^2+q^2)\\
  B^{V}_{S(V)S(V)}&=&\frac{2(m^2-q^2)(m^2-2mM-q^2)}{3M^2q^2}~.
\end{eqnarray}
Finally for the third term, which, being proportional to $q^\mu q^\nu$, 
is associated to a scalar quantity, as in Eq.~\eqref{eq:A862}, we get
\begin{eqnarray}
  \label{eq:V011}
    \tilde A^{S}_{S(V)S(V)}&=&\frac{Q_-^2Q_+}{3M^2q^4}\\
    B^{S}_{S(V)S(V)}&=&-\frac{(m^2-q^2)^2(m^2-2mM-q^2)}{3M^2q^4}~.
\end{eqnarray}
Again the symmetry relation
\begin{equation}
  \label{eq:V020}
  \Pi^{*(2)\mu\nu}_{S(A)S(A)}=-\Pi^{*(2)\mu\nu}_{S(V)S(V)}\bigm|_{m\to-m}
\end{equation}
holds.

Note that the interchange of the initial and final vertex entails an
interchange also of the indices $\mu$ and $\nu$.

\appendix
\section{The elementary functions $T^{*[m,n]}$}

\label{sec:AppD}

The functions defined in Eq.~\eqref{eq:A992} are given by
\begin{eqnarray}
  \label{eq:AX005}
  T^{*[1,0]}_{S+}&=&2\T^{[0]}\\
  T^{*[2,0]}_{S+}&=&2q^2\rho\T^{[0]}\\
  T^{*[2,0]}_{S-}&=&4q_0\T^{[1]}\\
  T^{*[3,0]}_{S+}&=&-\frac{2}{3}(4m^2\q^2-3\rho^2q^4)\T^{[0]}
  +\frac{8}{3}(3q_0^2+\q^2)\T^{[2]}\\
  T^{*[4,0]}_{S+}&=&2\rho q^2(\rho^2q^4-4m^2\q^2)\T^{[0]}
  +8\rho q^2(3q_0^2+\q^2)\T^{[2]}\\
  T^{*[5,0]}_{S+}&=&2\left[\left(q^4\rho^2-4m^2\q^2\right)^2
    -\frac{64}{5}m^4\q^4\right]\T^{[0]}\\
  \nonumber
  &+&\left[16q^4\rho^2(3q_0^2+\q^2)-\frac{64}{5}m^2\q^2
    (5q_0^2+\q^2)\right]\T^{[2]}\\
  \nonumber
  &+&32\left[\left(q_0^2+\q^2\right)^2-\frac{4}{5}\q^4\right]
  \T^{[4]}\\
  T^{*[1,0]}_{V+}&=&-2\frac{\q^2}{q^2}\T^{[1]}\\
  T^{*[2,0]}_{V+}&=&-2\rho\q^2\T^{[1]}\\
  T^{*[2,0]}_{V-}&=&\frac{4q_0\q^2}{3q^2}\left(m^2\T^{[0]}-4\T^{[2]}\right)\\
  T^{*[3,0]}_{V-}&=&2\rho\q^2T^{*[2,0]}_{V-}\\
  T^{*[4,0]}_{V-}&=&\frac{4m^2q_0\q^2}{q^2}\left(q^4\rho^2-\frac{4}{5}m^2\q^2
    \right)\T^{[0]}\\
  &-&\frac{16q_0\q^2}{q^2}\left[q^4\rho^2-m^2\left(q_0^2+\frac{7}{5}
      \q^2\right)\right]\T^{[2]}
  \nonumber\\
  &-&\frac{32q_0\q^2}{q^2}\left[q_0^2+\frac{3}{5}\q^2\right]\T{[4]}\nonumber\\
  T^{*[1,0]}_{L+}&=&-\frac{2\q^2}{3q^4}\left[m^2q_0^2\T^{[0]}-
    (q^2+4\q^2)\T^{[2]}\right]\\
  T^{*[2,0]}_{L+}&=&q^2\rho T^{*[1,0]}_{L+}\\
  T^{*[3,0]}_{L+}&=&-\frac{2}{3}\frac{m^2q_0^2\q^2}{q^4}
  \left[q^4\rho^2-\frac{12}{5}m^2\q^2\right]
  \T^{[0]}\\
  \nonumber&+&\frac{2}{3}\frac{\q^2}{q^4}
  \left[(q_0^2+3\q^2)q^4\rho^2-\frac{4}{5}m^2
    (q_0^2+5\q^2)(\q^2+5q_0^2)\right]\T^{[2]}\\
  \nonumber&+&\frac{8}{3}\frac{\q^2}{q^4}\left[q^4+\frac{48}{5}q_0^2q^2
    \right]\T{[4]}\\
  T^{*[1,0]}_{T+}&=&-\frac{4}{3}m^2\T^{[0]}+\frac{4}{3}\T^{[2]}\\
  T^{*[2,0]}_{T+}&=&q^2\rho T^{*[1,0]}_{T+}\\
  T^{*[3,0]}_{T+}&=&-\frac{4}{3}m^2\left[q^4\rho^2-\frac{4}{5}m^2
      \q^2\right]\T^{[0]}\\
  \nonumber&+&\frac{4}{3}\left[q^4\rho^2-4m^2\left(q_0^2+\frac{2}{5}\q^2\right)
  \right]\T^{[2]}\\
  \nonumber&+&\frac{16}{3}\left[q_0^2+\frac{1}{5}\q^2\right]\T^{[4]}~.\\
\end{eqnarray}

\section{The elementary functions $U^{*[n]}$}

\label{sec:appA}

In Sec.~\ref{sec:4} we have derived the functions $U^{*[0]}_{rel}$. 
Actually we also need $U^{*[1]}_{rel}$ and $U^{*[2]}_{rel}$, which be be 
provided in this appendix.
It is clear from the derivation of $U^{*[0]}_{rel}$
that the entire class of functions defined by Eq.~\eqref{eq:C009} has a 
common structure, namely
\begin{equation}
  \label{eq:A331}
  U^{*[n]}_{rel}=\alpha_0^{[n]}+\alpha_1^{[n]}l_1
  +\alpha_2^{[n]}l_2+\alpha_3^{[n]}l_3~.
\end{equation}
The coefficients are given by
\begin{subeqnarray}
  \alpha_0^{[0]}&=&0\\
  \alpha_1^{[0]}&=&-\frac{\rho}{16\pi^2}\\
  \alpha_2^{[0]}&=&\frac{q_0\rho+2E_F}{32\pi^2\q}\\
  \alpha_3^{[0]}&=&-\frac{\Delta^*}{64\pi^2}\\
  \alpha_0^{[1]}&=&-\dfrac{\rho k_F}{32\pi^2}\\
  \alpha_1^{[1]}
  &=&\frac{q_0}{32\pi^2}\left[\rho^2-\frac{2m^2}{q^2}\right]\\
  \alpha_2^{[1]}&=&\frac{1}{128\pi^2\q}\left[4k_F^2-
    (\q^2+q_0^2)\rho^2+\frac{4m^2q_0^2}{q^2}\right]
  \nonumber\\
~\\
  \alpha_3^{[1]}
  &=&\frac{q_0}{128\pi^2}\rho\Delta^*\\
  \alpha_0^{[2]}&=&\dfrac{k_Fq_0}{48\pi^2}\left[\rho^2-\frac{\rho E_F}{2q_0}
    -\dfrac{2m^2}{q^2}\right]\\
  \alpha_1^{[2]}&=&-\frac{\rho}{192\pi^2}\left[\rho^2(\q^2+3q_0^2)-
    \frac{6m^2}{q^2}(\q^2+q_0^2)\right]\\
  \alpha_2^{[2]}&=&\frac{1}{384\pi^2\q}\left\{\rho q_0\left[
      (3\q^2+q_0^2)\rho^2-\frac{12m^2\q^2}{q^2}\right]+8E_F^3\right\}\\
  \alpha_3^{[2]}&=&\frac{\Delta^*}{768\pi^2}\left[\frac{4m^2\q^2}{q^2}
    -(\q^2+3q_0^2)\rho^2\right]
\end{subeqnarray}

\section{The contact terms for the 2-indices $\Delta$-$N$ functions}
\label{sec:appE}

In this Appendix we list the 69 non-vanishing contact terms associated
to the two-indices LFs of Sec.~\ref{sec:3.4}.
We use the notation $\lambda^\alpha_{mn X_1(Y_1) X_2(Y_2)}$,
where $\alpha=S,V,LT$ is associated to the Lorentz components,
$X_i=M,E,C,S$  to the magnetic, electric, Coulomb or 
scalar nature of the two currents $(i=1,2)$ and $Y_i=V,A$ to the their vector or 
axial  parts, respectively.
They are given by:

\begin{align}
  \intertext{1) vertices $M(V)M(V)$}
  \lambda^S_{10M(V)M(V)}&=-\frac{3M_T^2}{4m^2q^2}
  \left(1-4\frac{M^2-q^2}{Q_+}\right)\\
  \lambda^S_{20M(V)M(V)}&=-\frac{3M_T^2}{4m^2q^2Q_+^2}\left(Q_--
    4(M^2-q^2)\right)\\
  \lambda^S_{30M(V)M(V)}&=-\frac{3M_T^2}{4m^2q^2Q_+^2}\\
  \lambda^{LT}_{10M(V)M(V)}&=\frac{3M_T^2q^2}{m^2Q_+^2}\\
  \intertext{2) vertices $M(V)E(V)$}
  \lambda^S_{10M(V)E(V)}&=\frac{3M_T^2}{4m^2q^2}
  \left(1-\frac{4m^2}{Q_+}\right)\\
  \lambda^S_{20M(V)E(V)}&=\frac{3M_T^2}{4m^2q^2Q_+^2Q_-}\left(3q^4\rho^2-
    4mMq^2\rho-4m^2M_T\delta m\right)\\
  \lambda^S_{30M(V)E(V)}&=\frac{3M_T^2}{4m^2q^2Q_+^2Q_-}
  \left[3Q_+-4m(2m+M)\right]\\
  \lambda^S_{40M(V)E(V)}&=\frac{3M_T^2}{4m^2q^2Q_+^2Q_-}
  \\
  \lambda^{LT}_{10M(V)E(V)}&=\frac{3M_T^2q^2}{m^2Q_+^2Q_-}
  \left(Q_+-4m^2\right)\\
  \lambda^{LT}_{20M(V)E(V)}&=\frac{3M_T^2q^2}{m^2Q_+^2Q_-}\\
  \intertext{3) vertices $E(V)E(V)$}
  \lambda^S_{10E(V)E(V)}&=\frac{3M_T^2}{4m^2q^2Q_+}\left[15Q_+
    -4m(5m+6M)\right]\\
  \lambda^S_{20E(V)E(V)}&=\frac{3M_T^2}{4m^2q^2Q_+^2Q_-}
  \left[31q^4\rho^2-28m(2m+M)q^2\rho-20m^2M_T\delta m\right]\\
  \lambda^S_{30E(V)E(V)}&=\frac{3M_T^2}{4m^2q^2Q_+^2Q_-^2}
  [33q^4\rho^2\\&
  -4m(13m-5M)q^2\rho-4m^2(9m+7M)\delta m]\nonumber\\
  \lambda^S_{40E(V)E(V)}&=\frac{3M_T^2}{2m^2q^2Q_+^2Q_-^2}(9Q_-+16m\delta m)\\
  \lambda^S_{50E(V)E(V)}&=\frac{3M_T^2}{m^2q^2Q_+^2Q_-^2}\\
  \lambda^{LT}_{10E(V)E(V)}&=\frac{3M_T^2q^2}{m^2Q_+^2Q_-}[9Q_--8m(m-3M)]\\
  \lambda^{LT}_{20E(V)E(V)}&=\frac{6M_T^2q^2}{m^2Q_+^2Q_-^2}(5Q_-+8m\delta m)\\
  \lambda^{LT}_{30E(V)E(V)}&=\frac{12M_T^2q^2}{m^2Q_+^2Q_-^2}\\
  \intertext{4) vertices $C(V)C(V)$}
  \lambda^{LT}_{10C(V)C(V)}&=\frac{3M_T^2q^4}{m^2Q_+^2Q_-^2}
  (Q_+-3M^2+4q^2)\\
  \lambda^{LT}_{20C(V)C(V)}&=-\frac{3M_T^2q^4}{m^2Q_+^2Q_-^2}\\
  \intertext{5) vertices $M(V)M(A)$}
  \lambda^V_{20M(V)M(A)}&=-\frac{3iM_T\delta m }{m^2Q_+^2Q_-}q^2\rho\\
  \lambda^V_{30M(V)M(A)}&=\frac{3iM_T\delta m}{2m^2Q_+^2Q_-}\\
  \intertext{6) vertices $M(V)E(A)$}
  \lambda^V_{20M(V)E(A)}&=-\frac{3iM_T\delta m}{2m^2Q_+^2Q_-}\left[3Q_-+
    4m(2M-m)\right]\\
  \lambda^V_{30M(V)E(A)}&=-\frac{3iM_T\delta m}{2m^2Q_+^2Q_-}\\
  \intertext{7) vertices $E(V)M(A)$}
  \lambda^V_{20E(V)M(A)}&=\frac{6iM_T\delta m}{m^2Q_+^2Q_-^2}\\
  &\times\left\{-2q^4\rho^2+m(5m+3M)q^2\rho+2m^2q^2\right\}
  \nonumber\\
  \lambda^V_{30E(V)M(A)}&=-\frac{3iM_T\delta m}{2m^2Q_+^2Q_-^2}
  \left[Q_++4m(2m+M)\right]\\
  \lambda^V_{40E(V)M(A)}&=\frac{3iM_T\delta m}{2m^2Q_+^2Q_-^2}\\
  \intertext{8) vertices $E(V)E(A)$}
  \lambda^V_{20E(V)E(A)}&=-\frac{3iM_T\delta m}{2m^2Q_+^2Q_-^2}\left[
    3q^4\rho^2-4m\delta m q^2\rho-4m^2q^2\right]\\
  \lambda^V_{30E(V)E(A)}&=-\frac{3iM_T\delta m}{2m^2Q_+^2Q_-^2}\left\{
    3Q_+-4m(2m+M)\right\}\\
  \lambda^V_{40E(V)E(A)}&=-\frac{3iM_T\delta m}{2m^2Q_+^2Q_-^2}\\
  \intertext{9) vertices $M(A)M(A)$}
  \lambda^S_{10M(A)M(A)}&=-\frac{3\delta m^2}{4m^2q^2Q_-}\left[9Q_--8m(m-3M)
  \right]\\
  \lambda^S_{20M(A)M(A)}&=\frac{3\delta m^2}{2m^2q^2Q_+Q_-^2}
  \left\{-5q^4\rho^2+4m(3m+M)q^2\rho+4m^2q^2\right\}\\
  \lambda^S_{30M(A)M(A)}&=-\frac{3\delta m^2}{2m^2q^2Q_+^2Q_-^2}\\  
  &\times\left\{3q^4\rho^2-4m(2m+M)q^2\rho+4m^2M_T(3m+M)\right\}\\
  \lambda^S_{40M(A)M(A)}&=-\frac{3\delta m^2}{4m^2q^2Q_+^2Q_-^2}
  \left\{3Q_+-8mM_T\right\}\\
  \lambda^S_{50M(A)M(A)}&=-\frac{3\delta m^2}{4m^2Q_+^2Q_-^2}\\
  \lambda^{LT}_{10M(A)M(A)}&=-\frac{3\delta m^2q^2}{m^2Q_+^2Q_-^2}\\
  &\times\left\{3q^4\rho^2-4m(4m+3M)q^2\rho
    +4m^2M_T(5m+3M)\right\}\\
  \lambda^{LT}_{20M(A)M(A)}&=-\frac{3\delta m^2q^2}{m^2Q_+^2Q_-^2}
  \left\{Q_--4m^2\right\}\\
  \lambda^{LT}_{30M(A)M(A)}&=-\frac{3\delta m^2q^2}{m^2Q_+^2Q_-^2}\\
  \intertext{10) vertices $M(A)E(A)$}
  \lambda^S_{10M(A)E(A)}&=\frac{3\delta m^2}{4m^2q^2}\\
  \lambda^S_{20M(A)E(A)}&=-\frac{3\delta m^2\rho}{m^2Q_+Q_-}\\
  \lambda^S_{30M(A)E(A)}&=\frac{3\delta m^2}{2m^2Q_+^2Q_-^2}\left\{
    3q^2\rho^2-4m^2\right\}\\
  \lambda^S_{40M(A)E(A)}&=-\frac{3\delta m^2\rho}{m^2Q_+^2Q_-^2}\\
  \lambda^S_{50M(A)E(A)}&=\frac{3\delta m^2}{4m^2q^2Q_+^2Q_-^2}\\
  \lambda^{LT}_{10M(A)E(A)}&=\frac{3\delta m^2q^2}{m^2Q_+Q_-}\\
  \lambda^{LT}_{20M(A)E(A)}&=-\frac{6\delta m^2q^4\rho}{m^2Q_+^2Q_-^2}\\
  \lambda^{LT}_{30M(A)E(A)}&=\frac{3\delta m^2q^2}{m^2Q_+^2Q_-^2}\\
  \intertext{11) vertices $E(A)E(A)$}
  \lambda^S_{10E(A)E(A)}&=-\frac{3\delta m^2}{4m^2q^2Q_-}
  \left[5Q_-+8m\delta m\right]\\
  \lambda^S_{20E(A)E(A)}&=-\frac{3\delta m^2}{2m^2q^2Q_+Q_-^2}
  \left[5q^4\rho^2-4m(2m-M)q^2\rho-4m^2M_T\delta m\right]\\
  \lambda^S_{30E(A)E(A)}&=-\frac{3\delta m^2}{2m^2q^2Q_+^2Q_-^2}
  \left[5q^4\rho^2-4m(2m+M)q^2\rho-4m^2M_T\delta m\right]\\
  \lambda^S_{40E(A)E(A)}&=-\frac{3\delta m^2}{4m^2q^2Q_+^2Q_-^2}
  \left[5Q_--4m(2m-3M)\right]\\
  \lambda^S_{50E(A)E(A)}&=-\frac{3\delta m^2}{4m^2q^2Q_-^2Q_+^2}\\
  \lambda^{LT}_{10E(A)E(A)}&=-\frac{3\delta m^2q^2}{m^2Q_+Q_-^2}
  \left[3Q_-+4m\delta m\right]\\
  \lambda^{LT}_{20E(A)E(A)}&=-\frac{3\delta m^2 q^2}{m^2Q_-^2Q_+^2}
  \left[3Q_+-4mM_T\right]\\
  \lambda^{LT}_{30E(A)E(A)}&=-\frac{3\delta m^2 q^2}{m^2Q_-^2Q_+^2}\\
  \intertext{12) vertices $C(A)C(A)$}
  \lambda^{LT}_{10C(A)C(A)}&=\frac{3\delta m^2 q^4}{m^2Q_-^2Q_+^2}
  \left[3Q_+-(2m+M)^2\right]\\
  \lambda^{LT}_{20C(A)C(A)}&\frac{3\delta m^2 q^4}{m^2Q_-^2Q_+^2}\\
  \intertext{13) vertices $M(V)S(V)$}
  \lambda^S_{10M(V)S(V)}&=-\frac{iM_T \rho}{m Q_+}\\
  \lambda^S_{20M(V)S(V)}&=\frac{iM_T}{2m q^2Q_+}\\
  \intertext{14) vertices $E(V)S(V)$}
  \lambda^S_{10E(V)S(V)}&=\frac{iM_T}{m q^2Q_+}\left[Q_+-6mM_T\right]\\
  \lambda^S_{20E(V)S(V)}&=\frac{iM_T}{2m q^2 Q_+Q_-}
  \left[11Q_-+16m\delta m\right]\\
  \lambda^S_{30E(V)S(V)}&=\frac{2iM_T}{m q^2Q_+Q_-}\\
  \lambda^{LT}_{10E(V)S(V)}&=\frac{8iM_Tq^2}{m Q_+Q_-}
  \intertext{15) vertices $C(V)S(V)$}
  \lambda^{LT}_{10C(V)S(V)}&=-\frac{2iM_Tq^2}{m Q_+Q_-}\\
  \lambda^{V}_{20C(V)S(V)}&=-\frac{iM_T}{m Q_+Q_-}
  \intertext{16) vertices $M(A)S(V)$}
  \lambda^{V}_{20M(A)S(V)}&=\frac{\delta m}{m Q_+Q_-}
  \intertext{17) vertices $E(A)S(V)$}
  \lambda^{V}_{20E(A)S(V)}&=-\frac{\delta m}{m Q_+Q_-}
  \intertext{18) vertices $S(V)S(V)$}
  \lambda^{S}_{10S(V)S(V)}&=-\frac{4}{3q^2}\\
  \tilde\lambda^{S}_{10S(V)S(V)}&=-\frac{m^2+2mM-2M^2+3q^2}{3q^4}\\
  \tilde\lambda^{S}_{10S(V)S(V)}&=\frac{1}{3q^4}
  \intertext{19) vertices $E(V)S(A)$}
  \lambda^V_{20E(V)S(A)}&=\frac{2M_T}{mQ_+Q_-}
  \intertext{20) vertices $M(A)S(A)$}
  \lambda^S_{10M(A)S(A)}&=i\frac{\delta m}{mq^2}\\
  \lambda^S_{20M(A)S(A)}&=-\frac{2i\delta m\rho}{mQ_+Q_-}\\
  \lambda^S_{30M(A)S(A)}&=\frac{i\delta m}{mq^2Q_+Q_-}\\
  \lambda^{LT}_{10M(A)S(A)}&=\frac{4i\delta m q^2}{mQ_+Q_-}
  \intertext{21) vertices $E(A)S(A)$}
  \lambda^S_{10E(A)S(A)}&=-i\frac{\delta m}{mq^2Q_-}(3Q_-+4m\delta m)\\
  \lambda^S_{20E(A)S(A)}&=-i\frac{\delta m}{mq^2Q_+Q_-}
  \left[3Q_-+4m(2M-m)\right]\\
  \lambda^S_{30E(A)S(A)}&=-i\frac{\delta m}{mq^2Q_+Q_-}\\
  \lambda^{LT}_{10E(A)S(A)}&=-4i\frac{\delta m q^2}{mQ_+Q_-}~.
\end{align}

\end{document}